\documentclass[twocolumn,showpacs,preprintnumbers,superscriptaddress,amsthm,amsmath,amssymb]{revtex4}
\usepackage{graphicx}
\usepackage{dcolumn}
\usepackage{bm}
\usepackage{color}
\usepackage{multirow}
\usepackage{CJK}
\usepackage{ltxtable}
\usepackage{rotating}
\usepackage{array}
\usepackage{booktabs,longtable}
\usepackage[colorlinks,linkcolor=blue,anchorcolor=blue,citecolor=red]{hyperref}
\usepackage{setspace}

\begin{document}
\begin{CJK*}{GBK}{song}

\title{Probing spin and pseudospin symmetries in deformed nuclei \\by Green's function method}

\author{Ting-Ting Sun}
\email{ttsunphy@zzu.edu.cn}
\affiliation{School of Physics and
Microelectronics, Zhengzhou University, Zhengzhou 450001, China}
\affiliation{Guangxi Key Laboratory of Nuclear Physics and Nuclear Technology,
Guangxi Normal University, Guilin 541004, China}

\author{Bing-Xin Li}
\affiliation{School of Physics and Microelectronics, Zhengzhou University,
Zhengzhou 450001, China}

\author{Kun Liu}
\affiliation{School of Physics and Microelectronics, Zhengzhou University,
Zhengzhou 450001, China}

\date{\today}

\begin{abstract}
Spin and pseudospin symmetries play vital roles in nuclear physics and have been studied extensively in spherical
nuclei. However, there are much fewer studies in deformed systems and there remain many open questions. In this
work, for the first time, we examine the possible spin and pseudospin
symmetries in deformed nuclei by solving a coupled-channel Dirac equation
with the Green's function method, which provides a novel way to exactly
determine the single-particle levels and properly describe the spacial
density distributions. Taking axially-deformed nucleus $^{154}$Dy as an
example, the spin doublets with a combination of Nilsson levels $\Lambda\pm
1/2[\mathcal{N},n_z, \Lambda]$ and pseudospin doublets with a combination of
$\widetilde{\Lambda}\pm 1/2[\widetilde{\mathcal{N},n_z, \Lambda}]$ are
determined. Different behaviors are displayed for the spin and pseudospin
doublets. For the spin partners, those with smaller angular momentum $l$ and
the third component $\Lambda$ owns better symmetry such as the $1p$ doublet
while good pseudospin symmetry appears in partners locating close to the
continuum threshold. By examining the single-particle Nilsson levels
$\Omega[\mathcal{N},n_z,\Lambda]$ and the energy splittings between the
partners, the conservation and breaking of SS and PSS are examined at
different deformations. In the prolate side, the Nilsson levels for the spin
and pseudospin doublets are almost parallel and the energy splittings are
stable against varying deformations. By examining the density distributions,
great similarities have been observed in the upper components for the spin
doublets while great similarities in the lower component for the pseudospin
doublets. Besides, these similarities maintain well at different
deformations.
\end{abstract}



\maketitle

\end{CJK*}

\section{Introduction}\label{Sec:Introduction}

Symmetries in the single-particle spectra of atomic nuclei are of great
significance to nuclear structure and have been extensively discussed in the
literature~\cite{Ginocchio_PhysRep_2005,Liang_PhysRep_2015,Leviatan_PLB_2001,Shen_PLB_2018}.
The breaking of spin symmetry~(SS), i.e., the remarkable spin-orbit splitting
for the spin doublets ($n,l,j=l \pm 1/2$) caused by the spin-orbit potential
(SOP), lays the foundation for the understanding of the traditional magic
numbers in nuclear physics~\cite{Otto_PhysRev_1949,Mayer_PhysRev_1949}.
Afterwards, based on the observation of the near-degeneracy in single-particle
levels with quantum numbers ($n,l,j=l+1/2$) and ($n-1,l+2,j=l+3/2$), a new
symmetry, namely the pseudospin symmetry~(PSS)~is introduced in nuclear
physics, and the two levels are viewed as pseudospin doublets denoted by the
pseudo quantum numbers ($\tilde{n}=n-1$, $\tilde{l}=l+1$, $j=\tilde{l}\pm
1/2$)~\cite{Hecht_NPA_1969,Arima_PLB_1969}. Significantly, PSS has been used
to explain a number of phenomena in nuclear structure such as nuclear
deformation~\cite{Bohr_PhysScr_1982}, superdeformation~\cite{Dudek_PRL_1987},
magnetic moment~\cite{Nazarewicz_PRL_1990}, and identical rotational
bands~\cite{Byrski_PRL_1990}.

In early years, comprehensive efforts have been made to understand the origin
of PSS. Apart from relabeling the quantum numbers, the explicit
transformations from a normal state ($l,~s$) to a pseudo state
($\tilde{l},~\tilde{s}$) have been proposed in
Refs.~\cite{Castanos_PLB_1992,Bahri_PRL_1992,Blokhin_PRL_1995}. In 1997,
Ginocchio made a substantial progress and clearly showed that PSS is a
relativistic symmetry in the Dirac Hamiltonian and becomes exactly conserved
when the attractive scalar and repulsive vector potentials are equal in size
and opposite in sign, i.e., $\Sigma(r) \equiv
S(r)+V(r)=0$~\cite{Ginocchio_PRL_1997}. He also claimed that the
pseudo-orbital angular momentum $\tilde{l}$ is nothing but the orbital angular
momentum of the lower component of the Dirac wave
function~\cite{Ginocchio_PRL_1997}. Meanwhile, the single-nucleon wave
functions of the lower component for the pseudospin doublets exist some
similarities~\cite{Ginocchio_PRC_1998}. However, the PSS is always broken in
real nuclear systems due to the nonzero potentials $\Sigma(r)$. Later, Meng
$\textit{et al}$. proposed a more general condition $d \Sigma(r)/d r=0$ for
PSS, which can be approximately satisfied in exotic nuclei with highly
diffused potentials~\cite{Meng_PRC_1998}. Besides, he also pointed out that
the extent of the conservation of PSS is connected with the competition
between the pseudocentrifugal barrier (PCB) and the pseudospin-orbit potential
(PSOP)~\cite{Meng_PRC_1999}. Afterwards, the SS and PSS have been studied
extensively, such as PSS and SS in
hypernuclei~\cite{Song_CPL_2009,Song_CPL_2010,Song_CPL_2011,Sun_PRC_2017,Lu_JPG_2017},
SS in anti-nucleon
spectra~\cite{Zhou_PRL_2003,Mishustin_PRC_2005,He_EPJA_2006,Liang_EPJA_2010},
PSS in the single-particle resonate
states~\cite{Guo_PRC_2005,Guo_PRC_2006,Zhang_CPL_2007,Lu_PRL_2012,Lu_PRC_2013,Liu_PRA_2013,Sun_PRC_2019_034310,ShiXX_PLB_2020,Liu_PLB_2022},
perturbative interpretation of SS and PSS~\cite{Liang_PRC_2011,Li_CPC_2011},
and PSS in supersymmetric quantum
mechanics~\cite{Liang_PRC_2013,Shen_PRC_2013}.

PSS has also been observed in deformed nuclei~\cite{Mottelson_NPA_1991}. In
axially deformed nuclei, two single-particle orbitals with asymptotic Nilsson
quantum numbers~($\Omega=\Lambda+1/2~[\mathcal{N},n_{3},\Lambda]$) and
($\Omega=\Lambda+3/2~[\mathcal{N},n_{3},\Lambda+2]$) are redefined as a
pseudospin doublet ($\widetilde{\Omega}=\widetilde{\Lambda}\pm 1/2
~[\widetilde{\mathcal{N}}=\mathcal{N}-1,\widetilde{n}_{3}={n}_{3},\widetilde{\Lambda}=\Lambda+1]$)
due to the quasidegeneracy between them~\cite{Bohr_PhysScr_1982}. Up to now, a
series of work has been done to study the SS and PSS in deformed
nuclei~\cite{Sugawara-Tanabe_PRC_1998,Sugawara-Tanabe_PRC_2002}. In
Ref.~\cite{Sugawara-Tanabe_PRC_2000}, based on the relativistic
mean-field~(RMF)~theory, Sugawara-Tanabe~\emph{et al}. pointed out for the
first time that the PSS is always hidden in the unnatural parts of the Dirac
wave function irrespective of the deformation. Later, they also found that
both SS and PSS exist in the deformed nuclei and respectively related with the
upper and lower components of the Dirac wave
functions~\cite{Sugawara-Tanabe_PRC_2002}. In Ref.~\cite{Lalazissis_PRC_1998},
the quasidegenerate pseudospin doublets are confirmed to exist near the Fermi
surface for deformed nuclei by carrying out constrained deformed RMF
calculations. In Ref.~\cite{PRC2004Ginocchio_69_034303}, the Dirac
eigenfunctions in the RMF calculations of deformed nuclei were examined
extensively. In Ref.~\cite{Guo_PRL_2014}, based on the similarity
renormalization group theory, Guo~\emph{et al} investigated the pseudospin
symmetry in deformed nuclei and confirmed the crucial role of the
nonrelativistic term, the spin-orbit term, and the dynamical term. In
Ref.~\cite{LiuQ_PLB_2023}, based on the complex momentum
representation~(CMR)~method, the PSS in resonant states were explored in
deformed nuclei and found to be approximately reserved with small splittings
for energies, widths, and density distributions.

In recent years, Green's function (GF) method has achieved great successes in
describing continuum and has been widely applied in nuclear physics, such as
in the studies of the single-particle structures including the resonant
states~\cite{Sun_PRC_2014,Sun_PRC_2019_034310,TTSun_JPG_2016,SHRen_PRC_2017,Qu_PRC_2019,Qu_PRC_2022},
halos in exotic
nuclei~\cite{Oba_PRC_2009,Zhang_PRC_2011,TTSun_Sci_2016,Sun_PRC_2019_054316},
and collective
excitations~\cite{Matsuo_NPA_2001,Matsuo_PRC_2005,Matsuo_PRC_2010,Shimoyama_PRC_2011,Shimoyama_PRC_2013,Matsuo_PRC_2015}.
This method is very convenient to be used in combination with different
nuclear models. In 2014, we applied the Green's function method to the RMF
theory (RMF-GF) to investigate the single-particle resonant states for the
first time~\cite{Sun_PRC_2014}. In 2016, by further including the pairing
correlation, we applied Green's function method to the relativistic continuum
Hartree-Bogoliubov theory (RCHB-GF) to study the halo
phenomena~\cite{TTSun_Sci_2016}. In 2019, by including the blocking effects,
we extended the self-consistent Green's function continuum
Skyrme-Hartree-Fock-Bogoliubov theory~\cite{Zhang_PRC_2011} to the
descriptions of odd-$A$ nuclei~\cite{Sun_PRC_2019_054316,Huo_NST_2023}. In
2020, by further including the deformation, we applied the Green's function
method to solve a coupled-channel Dirac equation with axially quadrupole
deformed potentials and analyzed the deformed halo in
$^{37}$Mg~\cite{Sun_PRC_2020}. According those studies, Green's function
method has shown great advantages such as treating the single-particle bound
states and the continuum on the same footing, determining directly the
energies and widths for resonances, and describing properly the the spatial
density distributions. Recently, based on the Green's function method, a novel
way by searching for the extremes of the density of
states~\cite{Chen_CPC_2020} or the poles of Green's
function~\cite{Wang_NST_2021} has been proposed to determine the resonant
states, which can exactly determine the energies and widths for the bound and
resonant states regardless of the width. Besides, this method can describe the
resonant states in any potential without any requirement on the potential
shape. As an application, the conservation and breaking of PSS in the
single-nucleon resonant states have has been examined from the PSS limit to
finite depth potentials, and in the PSS limit, besides strictly the same
energies and widths between the PS partners, identical density distributions
of the lower component are found for the first time~\cite{PLB2023TTSun}.
Furthermore, a uniform description of pseudospin symmetry in bound and
resonant states has been given.

In this work, the possible SS and PSS of single-particle bound states in
deformed nuclei are examined by the Green's function method. The effects of
deformation on these symmetries are discussed. The paper is organized as
follows. The theoretical framework of the Green's function method for solving
the coupled-channel Dirac equation is presented in Sec.~\ref{Sec:Theory}.
After the numerical details in Sec.~\ref{Sec:NumDetails},
Sec.~\ref{Sec:Results} is devoted to the discussions of the numerical results,
where the conservation and breaking of the SS and PSS in deformed nuclei are
illustrated by analyzing the energy splittings and the density distributions.
Finally, a summary is given in Sec.~\ref{Sec:Summary}.

\section{Theoretical Framework}
\label{Sec:Theory}

To explore the SS and PSS in deformed nuclei with the RMF theory, the Dirac
equations governing the motion of the nucleons will be examined,
\begin{equation}
 \lbrace {\bm \alpha}\cdot {\bm p}+ V(\bm r) + \beta[M + S(\bm r)]\rbrace \psi(\bm r)=\varepsilon\psi(\bm r),
\label{Eq:Dirac}
\end{equation}
where ${\bm \alpha}$ and $\beta$ are Dirac matrices, $M$ is the mass of
nucleon, and $S(\bm r)$ and $V(\bm r)$ are the scalar and vector potentials,
respectively, which are adopted as the axially quadrupole-deformed potentials,
\begin{subequations}
  \begin{eqnarray}
   S(\bm r) = S_0(r) + S_2(r)Y_{20}(\theta,\phi), \\
   V(\bm r) = V_0(r) + V_2(r)Y_{20}(\theta,\phi),%
  \end{eqnarray}%
  \label{Eq:potential}%
\end{subequations}%
with $S_{0}(r)$ and $V_{0}(r)$ being the spherical components while
$S_{2}(r)$$Y_{20}$($\theta$,$\phi$) and $V_{2}(r)$$Y_{20}$($\theta$,$\phi$)
the quadrupole parts.

For a nucleon in an axially quadrupole-deformed potential, parity $\pi$ and
the $z$ component $\Omega$ of the angular momentum are good quantum numbers,
and the single-particle wave function can be expanded in terms of spherical
Dirac spinors,
\begin{equation}
  \psi_{\Omega} = \sum_{\kappa}
  \left(\begin{array}{c}
    i \frac{G_{\Omega \kappa}(r)}{r} \\[3pt]
    \frac{F_{\Omega \kappa}(r)}{r} \sigma \cdot \hat{\boldsymbol{r}} \\
    \end{array}
  \right)
  Y_{\kappa \Omega}(\theta, \phi),
\end{equation}
where $G_{\Omega\kappa}(r)/r$ and $F_{\Omega\kappa}(r)/r$ are, respectively,
the radial wave function for the upper and lower components,
$Y_{\kappa\Omega}(\theta,\phi)$ are the spinor spherical harmonics, and the
quantum number $\kappa$ is related with the orbital angular momentums $l$ and
the total angular momentums $j$,
\begin{equation}
  \begin{cases}
   l = \kappa, j = \kappa - \frac{1}{2}, & \text{if  } \kappa > 0,\\
   l = -\kappa - 1, j = -\kappa - \frac{1}{2}, & \text{if  } \kappa < 0,
  \end{cases}
\end{equation}
which labels different spherical partial waves or ``channels''.

Then the Dirac equation~(\ref{Eq:Dirac}) is transformed into a coupled-channel
form of radial wave functions,
\begin{subequations}
  \begin{eqnarray}
   &&0 = \frac{dG_{\Omega \kappa}}{dr} + \frac{\kappa}{r} G_{\Omega \kappa} - (\varepsilon_{\Omega} + 2M)F_{\Omega \kappa} \nonumber\\[4pt]
   &&~~~~+ \sum_{\kappa' \lambda}(V_{\lambda} - S_{\lambda})A(\lambda,\kappa',\kappa,\Omega)F_{\Omega \kappa'}, \\[4pt]
   &&0 = \frac{dF_{\Omega \kappa}}{dr} - \frac{\kappa}{r} F_{\Omega \kappa} + \varepsilon_{\Omega} G_{\Omega \kappa} \nonumber\\[4pt]
   &&~~~~- \sum_{\kappa' \lambda}(V_{\lambda} + S_{\lambda})A(\lambda,\kappa',\kappa,\Omega)G_{\Omega \kappa'},%
  \end{eqnarray}%
  \label{Eq:CCDirac}%
\end{subequations}%
where the couplings among different spherical channels are governed by the
deformed potentials,
\begin{equation}
  \upsilon_{\kappa \kappa'}^\pm = \sum_{\lambda}(V_{\lambda} \pm S_{\lambda})A(\lambda,\kappa',\kappa,\Omega),
\end{equation}
in which the index $\lambda=0$ and $2$, respectively, for the spherical and
quadrupole parts of the potentials, and $A(\lambda,\kappa',\kappa,\Omega)$ can
be expressed as,
\begin{eqnarray}
 &&A(\lambda,\kappa',\kappa,\Omega)  \\
 &&~~= \langle Y_{\kappa \Omega} \vert Y_{\lambda 0} \vert Y_{\kappa' \Omega} \rangle \nonumber \\
 &&~~= {(-1)}^{\Omega+\frac{1}{2}} \frac{\hat{j} \hat{j'}}{\sqrt{4 \pi}}
 \left(\begin{array}{ccc}
 j & ~\lambda~ & j' \\
 -\Omega & ~0~ & \Omega \\
 \end{array}\right)
 \left(\begin{array}{ccc}
 j' & ~~\lambda~ & j \\
 \frac{1}{2} & ~~0~ & -\frac{1}{2} \\
 \end{array}\right),\nonumber
\end{eqnarray}
with $\hat{j} = \sqrt{2j + 1}$. In the practical calculations, we must
truncate the partialwave expansion and use $N$ to represent the number of
spherical partial waves to be included for a given block $\Omega^\pi$.
Besides, note that the single-particle energy in Eq.~(\ref{Eq:CCDirac}) is
shifted by $M$ with respect to that in Eq.~(\ref{Eq:Dirac}).

Applying the Green's function method to solve the coupled-channel Dirac
Eq.~(\ref{Eq:CCDirac}), a Green's function will be constructed which is
defined as,
\begin{equation}
\lbrack\varepsilon - \hat{h}(\bm r)\rbrack\mathcal{G}({\bm r},{\bm r'};\varepsilon) = \delta({\bm r} - {\bm r'}),
\label{Eq:GF_def}
\end{equation}
where $\hat{h}({\bm r})$ is the Dirac Hamiltonian and $\varepsilon$ is an
arbitrary single-particle energy. In the axially quadrupole-deformed
potential, using the partial-wave expansion, the Green's function with a given
$\Omega$ can be expanded as,
\begin{equation}
  \mathcal{G}_{\Omega}({\bm r},{\bm r'};\varepsilon) = \sum_{\kappa \kappa'}Y_{\kappa \Omega}(\theta,\phi)
  \frac{\mathcal{G}_{\Omega \kappa \kappa'}({r},{r'};\varepsilon)}{rr'}Y_{\kappa' \Omega}^{*}(\theta', \phi'),
  \label{Eq:GFexp}
\end{equation}
where $\mathcal{G}_{\Omega \kappa \kappa'}({r},{r'};\varepsilon)$ is the
radial Green's function coupling the partial waves $\kappa$ and $\kappa'$, and
it is in a $2N\times 2N$ matrix form,
\begin{equation}
   \mathcal{G}_{\Omega \kappa \kappa'}({r},{r'};\varepsilon) =
   \left(\begin{array}{cc}
   \mathcal{G}_{\Omega \kappa \kappa'}^{(11)}({r},{r'};\varepsilon){}{} & \mathcal{G}_{\Omega \kappa \kappa'}^{(12)}({r},{r'};\varepsilon) \\[4pt]
   \mathcal{G}_{\Omega \kappa \kappa'}^{(21)}({r},{r'};\varepsilon){}{} & \mathcal{G}_{\Omega \kappa \kappa'}^{(22)}({r},{r'};\varepsilon) \\
   \end{array}\right).
\end{equation}

According to the definition in Eq.~(\ref{Eq:GF_def}), the radial Green's
function $\mathcal{G}_{\Omega \kappa \kappa'}(r,r';\varepsilon)$ satisfies the
following coupled-channel equation,
\begin{eqnarray}
  && \left(\begin{array}{cc}
   -\varepsilon & -\frac{d}{dr} + \frac{\kappa}{r}\\[4pt]
   \frac{d}{dr} + \frac{\kappa}{r} & -\varepsilon - 2M\\
   \end{array}\right)
   \mathcal{G}_{\Omega \kappa \kappa'}({r},{r'};\varepsilon)\\[5pt]
   &&+ \sum_{\kappa''}
   \left(\begin{array}{cc}
   \nu_{\kappa \kappa''}^{+} & 0\\[4pt]
   0 & \nu_{\kappa \kappa''}^{-} \\
   \end{array}\right)
   \mathcal{G}_{\Omega \kappa'' \kappa'}(r,r';\varepsilon)=\frac{\delta(r - r')}{rr'}J, \nonumber
\end{eqnarray}
where
\begin{equation}
  J=
  \left(\begin{array}{cc}
  1{}{} & 0 \\
  0{}{} & -1 \\
  \end{array}\right)
  \otimes I_{N},
\end{equation}
with $I_N$ being the $N$-dimensional unit matrix.

For the single-particle bound states, the density of states~(DOS)
$n(\varepsilon)$ exhibits discrete $\delta$ functions and can be easily
expressed as,
\begin{equation}
  n(\varepsilon) = \sum_{n}\delta(\varepsilon - \varepsilon_{n}),
\end{equation}
where $\varepsilon$ is a real single-particle energy and $\varepsilon_{n}$
represents the eigenvalues of the Dirac equation. With an infinitesimal
imaginary part ``$i\epsilon$'' to the real energy $\varepsilon$,
$n(\varepsilon)$ can be calculated by integrating the imaginary part of the
Green's function in the coordinate space. For a given block $\Omega^\pi$, it
can be expressed as,
\begin{eqnarray}
 &&n_\Omega(\varepsilon)  \label{Eq:dos}\\
 &=& -\frac{2}{\pi}\sum_{k}\int dr{\rm Im}\lbrack \mathcal{G}_{\Omega \kappa \kappa}^{(11)}(r,r;\varepsilon + i\epsilon)
 + \mathcal{G}_{\Omega \kappa \kappa}^{(22)}(r,r;\varepsilon + i\epsilon)\rbrack.\nonumber
\end{eqnarray}
With the infinitesimal imaginary part ``$i\epsilon$'', the DoSs for discrete
single-particle states in forms of $\delta$ functions (no width) are simulated
by Lorentzian functions with the full width at half maximum (FWHM) of
2$\epsilon$.

For the constructions of the Green's function and other details in solving the
couple-channel Dirac equations with the Green's function method, it can be
referred to our previous work in Ref.~\cite{Sun_PRC_2020}.

\section{Numerical Details}
\label{Sec:NumDetails}
In the present study,
potentials in Woods-Saxon forms for the radial part of the quadrupole deformed
potentials in Eq.~(\ref{Eq:potential}) are adopted
as~\cite{PRC2004Hamamoto_69_041306,PRC2010ZPLi_81_034311},
\begin{eqnarray}
   S_0(r)&=&S_{\rm WS}f(r),\nonumber \\
   V_0(r)&=&V_{\rm WS}f(r),\nonumber \\
   S_2(r)&=&-\beta S_{\rm WS}k(r),\nonumber \\
   V_2(r)&=&-\beta V_{\rm WS}k(r),
\end{eqnarray}
with
\begin{equation}
 f(r) = \frac{1}{1+{\rm exp}(\frac{r-R}{a})}, ~~~{\rm and}~~~
 k(r) = r\frac{df(r)}{dr}.
\end{equation}

To study the behavior of SS and PSS in the single-particle spectra, we take
the nucleus $^{154}$Dy as an example, which has a stable axially deformation
of $\beta=0.24$~\cite{ADNDT2016}. The mean-field Woods-Saxon potentials are
adopted with the depths of the scalar and vector potentials $S_{\rm WS} =
-405.0$~MeV and $V_{\rm WS} = 350.0$~MeV, the radius $R = 6.81$~fm, and the
diffuseness $a=0.67$~fm following Ref.~\cite{Guo_PRL_2014}.

To solve the coupled-channel Dirac equation in the coordinate space, a space
size of $R_{\rm box} = 20$~fm and a mesh step of $0.1$~fm are taken. For
calculating the density of states $n_\Omega(\varepsilon)$, the infinitesimal
parameter $\epsilon$ in Eq.~(\ref{Eq:dos}) is taken as $1.0\times$
$10^{-6}$~MeV and the energy step $d\varepsilon$ is $1.0\times$ $10^{-3}$ MeV.
With those parameters, the accuracy of the obtained single-particle energies
can be up to $1.0$~keV. Furthermore, a higher degree of accuracy can be
achieved for energies if a smaller energy steps $d\varepsilon$ is taken.

\section{RESULTS AND DISCUSSION}
\label{Sec:Results}

\begin{figure}[t!]
  \includegraphics[width=0.45\textwidth]{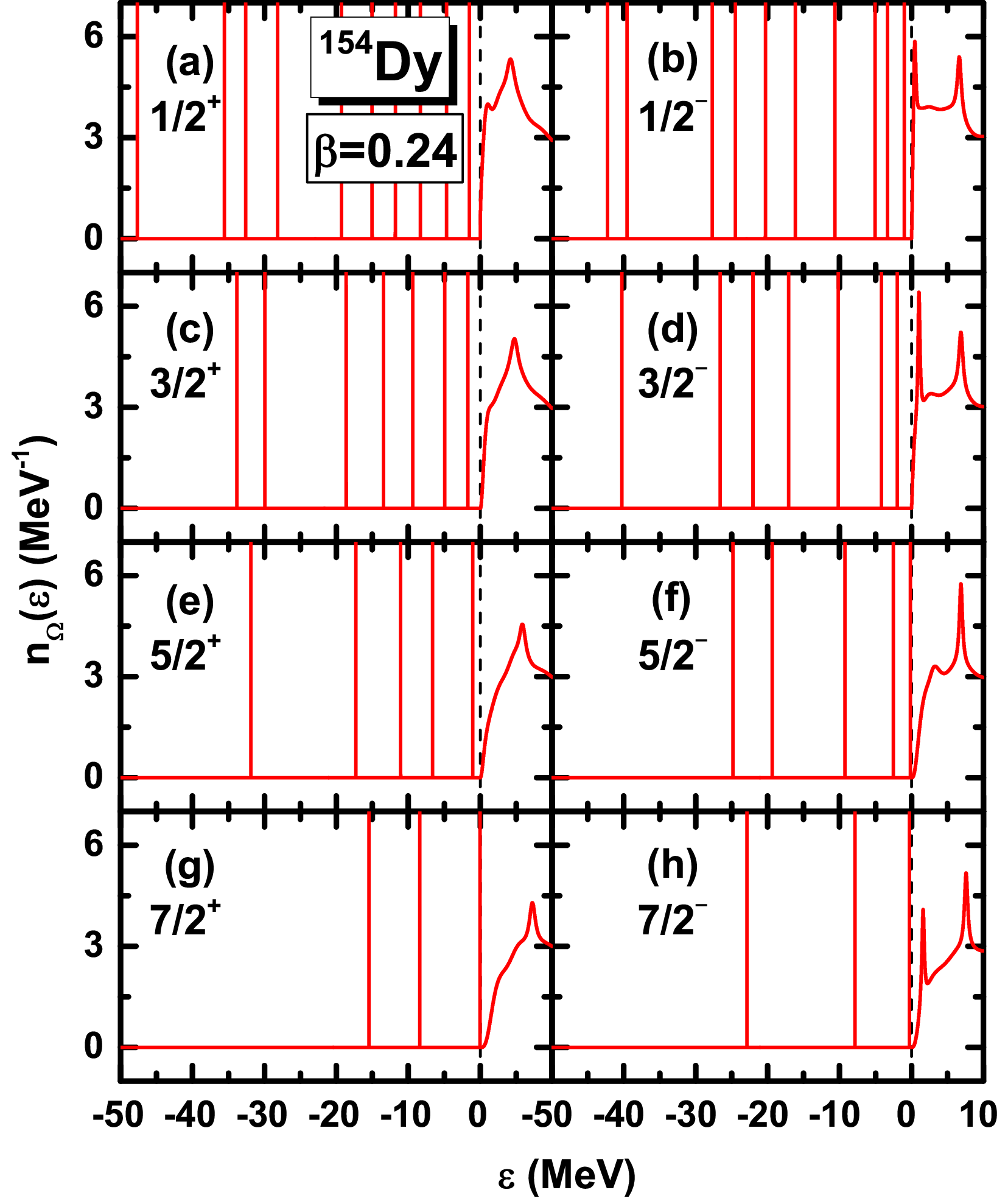}
\caption{Density of states $n_{\Omega}(\varepsilon)$ for nucleus $^{154}$Dy
with blocks $\Omega^\pi = {1/2}^\pm$, ${3/2}^\pm$, ${5/2}^\pm$, and
${7/2}^\pm$, obtained by solving the coupled-channel Dirac equations with
quadrupole-deformed Woods-Saxon potentials using the GF method. The
deformation parameter $\beta = 0.24$ and the number of coupled partial waves
$N=8$ are chosen. The dashed lines denote the continuum threshold.}
  \label{Fig1}
\end{figure}

With the Green's function method, the single-particle spectrum can be exactly
determined both for the bound and resonant states by searching for the poles
of Green's functions or extremes of density of states. In Fig.~\ref{Fig1}, the
density of states $n_{\Omega}(\varepsilon)$ for deformed nucleus $^{154}$Dy
with $\Omega^\pi$ =${1/2}^\pm$, ${3/2}^\pm$, ${5/2}^\pm$, and ${7/2}^\pm$ are
plotted as functions of single-particle energies $\varepsilon$ obtained by
solving the coupled-channel Dirac equation with the Green's function method,
where the deformation parameter $\beta=0.24$ and the number of coupled partial
waves $N=8$. The peaks in $\delta$-function shape below the continuum
threshold ($\varepsilon = 0$) correspond to the single-particle bound states
and the spectra with $\varepsilon > 0$ describe the continuum with peaks
therein being the single-particle resonant states. For the single-particle
bound states, the energies could be well determined directly by reading the
locations of the extremes of the DoSs. For the single-particle resonant
states, DoSs will be calculated on the complex energy plane
$\varepsilon=\varepsilon_r+i\varepsilon_i$ by scanning energies both along the
real $\varepsilon_r$ and imaginary $\varepsilon_i$ axes. For the details, it
can be referred to Refs.~\cite{Sun_PRC_2020,Chen_CPC_2020}.

\begin{table}[ht!]
  \center
\caption{Single-particle energies $\varepsilon$~(in MeV)~for the bound states
in $^{154}$Dy with $\Omega^\pi$ = ${1/2}^+$ and ${1/2}^-$ extracted from the
density of states shown in Fig.~\ref{Fig1}.}
  \label{Tab1}
  \begin{tabular}{ccccc}
  \hline\hline \\
  $\Omega[\mathcal{N}n_{z}\Lambda]$ &~~ $``+"$parity &~~~~& $\Omega[\mathcal{N}n_{z}\Lambda]$ &~~$``-"$parity \\[4pt]\hline
 1/2[000] & -47.681& &  1/2[110] & -42.268  \\[1pt]
 1/2[220] & -35.556& &  1/2[101] & -39.538  \\[1pt]
 1/2[211] & -32.605& &  1/2[330] & -27.708  \\[1pt]
 1/2[200] & -28.175& &  1/2[321] & -24.475  \\[1pt]
 1/2[440] & -19.295& &  1/2[310] & -20.314  \\[1pt]
 1/2[431] & -15.031& &  1/2[301] & -16.150  \\[1pt]
 1/2[420] & -11.810& &  1/2[550] & -10.661  \\[1pt]
 1/2[411] & -8.314 & &  1/2[541] & -5.057  \\[1pt]
 1/2[400] & -4.665 & &  1/2[530] & -3.311  \\[1pt]
          &        & &  1/2[510] & -1.008   \\
  \hline\hline
  \end{tabular}
\end{table}

In Table~\ref{Tab1}, taking the blocks $\Omega^\pi = {1/2}^+$ and ${1/2}^-$ as
examples, the single-particle energies $\varepsilon$ of the bound states
extracted from Fig.~\ref{Fig1} are listed, labeled by Nilsson quantum numbers
$\Omega[\mathcal{N},n_{z},\Lambda]$ with $\mathcal{N}$ being the principal
quantum number, $n_z$ the number of nodes of the wave functions in the $z$
direction, and $\Lambda$ the projection of the orbital angular momentum $l$
onto the $z$ axis. The nine positive parity states from top to bottom are
splitted from the spherical $1s_{1/2}$, $1d_{5/2}$, $1d_{3/2}$, $2s_{1/2}$,
$1g_{9/2}$, $1g_{7/2}$, $2d_{5/2}$, $2d_{3/2}$, and $3s_{1/2}$ states,
respectively. Similarly, the ten negative parity states are splitted from the
$1p_{3/2}$, $1p_{1/2}$, $1f_{7/2}$, $1f_{5/2}$, $2p_{3/2}$,~$2p_{1/2}$,
$1h_{11/2}$, $1h_{9/2}$, $2f_{7/2}$, and $3p_{3/2}$ states, respectively. As
shown in Table~\ref{Tab1}, energies for the bound states in a deformed Dirac
equation could be effectively obtained by the Green's function method. This
method can also strictly determine the energies and widths for resonant
states. However, in this work, we will not talk much for the resonant states
but mainly focus on the bound states.

\begin{figure}[t!]
  \includegraphics[width=0.45\textwidth]{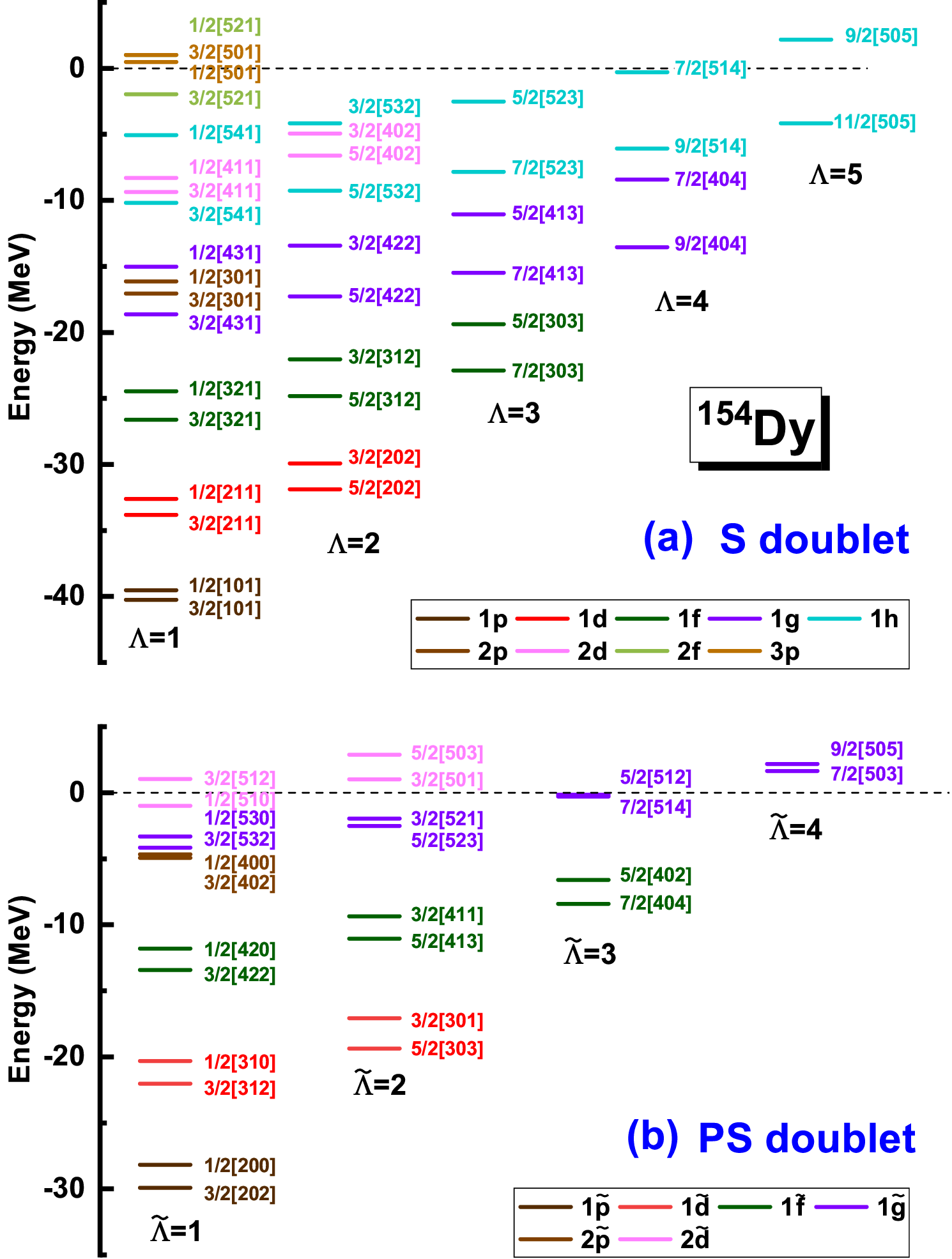}
\caption{Spin doublets $\Lambda\pm 1/2~[\mathcal{N}n_{z}\Lambda]$ and
pseudospin doublets $\widetilde{\Lambda}\pm
1/2~[\widetilde{\mathcal{N},n_{z},\Lambda}]$ in
$^{154}$Dy.}
  \label{Fig2}
\end{figure}

In the following, the SS and PSS in deformed nuclei are examined for the
single-particle bound states. In Fig.~\ref{Fig2}, taking $^{154}$Dy as an
example, we give the spin doublets as a combination of Nilsson levels with
quantum numbers $\Omega=\Lambda+1/2~[\mathcal{N},n_z,\Lambda]$ and
$\Omega=\Lambda-1/2~[\mathcal{N},n_z,\Lambda]$ and pseudospin doublets as a
combination of Nilsson levels with quantum numbers
$\Omega=\Lambda+1/2~[\mathcal{N},n_z,\Lambda]$ and
$\Omega=\Lambda+3/2~[\mathcal{N},n_z,\Lambda+2]$. The partners with the same
$\Lambda$ or $\widetilde{\Lambda}=\Lambda+1$ are put in the same column and
those with the same (pseudo) principal quantum number $n~(\widetilde{n})$ and
(pseudo) orbital angular momentum $l~(\widetilde{l})$ are denoted in the same
color. For spin doublets, the following points could be revealed: (i) For
partners with the same $\Lambda$, the SS becomes worse with increasing orbital
angular momentum $l$, showing greater energy differences between the partners.
In the case of $\Lambda=1$, the energy differences are respectively $0.73$,
$1.23$, $2.14$, $3.59$, and $5.14$~MeV for the $1p, 1d, 1f, 1g$, and $1h$ spin
doublets. (ii) For partners with the same $n$ and $l$, SS deteriorates with
larger $\Lambda$, e.g., the energy differences of the $1g$ doublets
corresponding to $\Lambda=1, 2, 3, 4$ are respectively $3.59$, $3.85$, $4.43$,
and $5.14$~MeV for the $(1/2[431], 3/2[431])$, $(3/2[422], 5/2[422])$,
$(5/2[413], 7/2[413])$, and $(7/2[404], 9/2[404])$ spin doublets. (iii) When
entering the continuum, an inversion of energy levels occurs for the spin
doublet with the spin down state being lower than the spin up state, such as
the partner $(1/2[501], 3/2[501])$.

\begin{figure}[tp!]
  \includegraphics[width=0.48\textwidth]{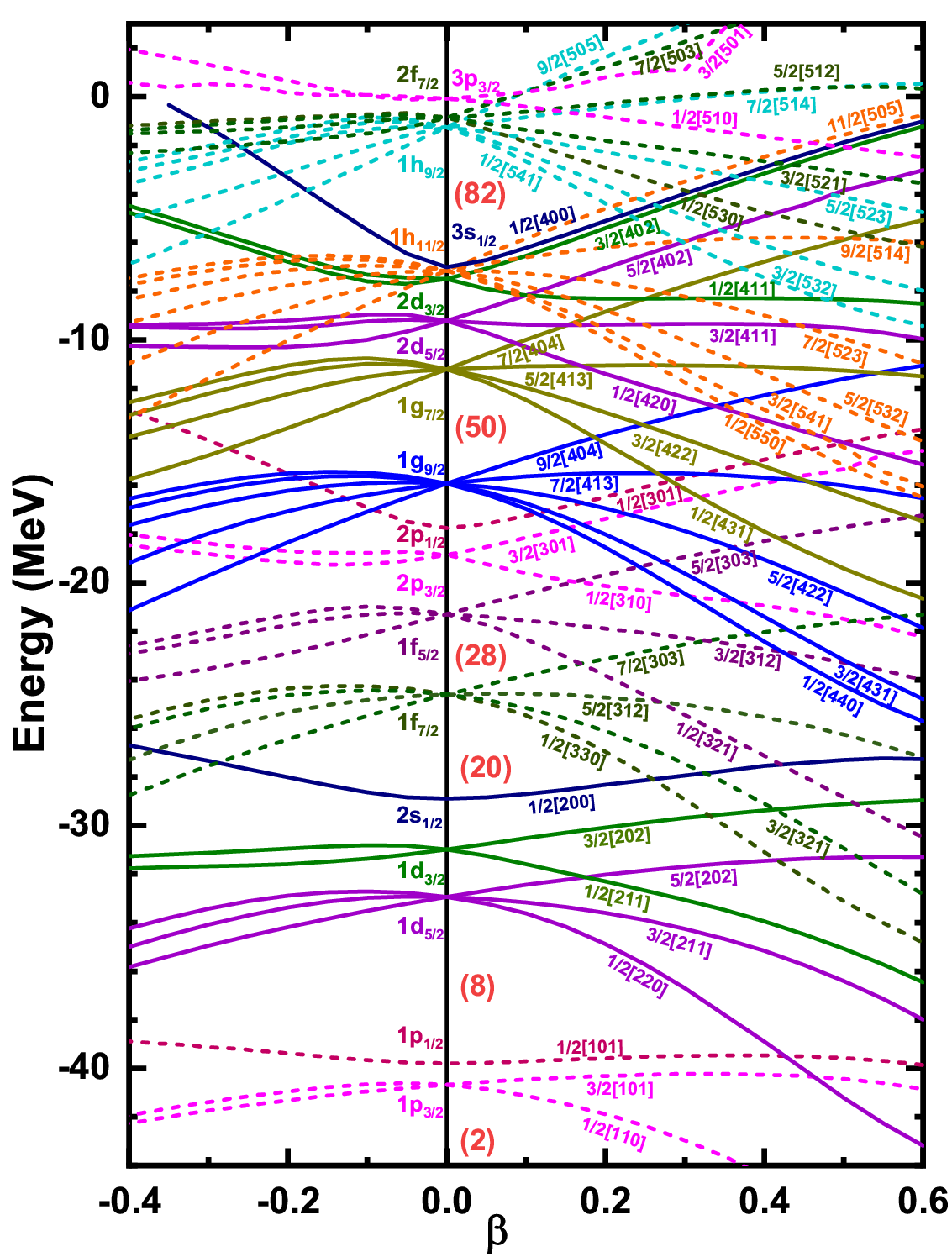}
\caption{Nilsson single-particle levels $\Omega~[\mathcal{N},n_{3},\Lambda]$
as a function of deformation parameter $\beta$ in quadrupole-deformed
Woods-Saxon potentials. The solid and dashed lines denote levels with positive and negative parities, respectively.}
  \label{Fig3}
\end{figure}

However, different behaviors are displayed for the pseudospin doublets: (i)
For partners with the same $\widetilde{\Lambda}$, the PSS is better maintained
with increasing pseudo-orbital angular momentum $\widetilde{l}$. In the case
of $\widetilde{\Lambda}=1$, the energy differences between the PS partners are
respectively $1.75$, $1.73$, $1.62$, and $0.86$~MeV for the $1\widetilde{p},
1\widetilde{d}, 1\widetilde{f}$, and $1\widetilde{g}$ PS doublets. (ii) For
partners with the same $\widetilde{n}$ and $\widetilde{l}$, the evolutions of
the PSS with $\widetilde{\Lambda}$ show different behaviors: for the
$1\widetilde{p}$ and $1\widetilde{d}$ PS partners, PSS becomes worse with
increasing $\widetilde{\Lambda}$ while it improves significantly for the
$1\widetilde{f}$ PS partners. (iii) There is an obvious threshold effect,
i.e., when the PS partner approaching to the continuum threshold, PSS becomes
much better and conserves approximately. (iv) Same as the spin doublet, when
entering the continuum, an inversion of energy levels occurs for the PS
doublet with the pseudospin down state being lower compared with the
pseudospin up state, such as the partners $(1/2[510], 3/2[512])$, $(3/2[501],
5/2[503])$, and $(7/2[503], 9/2[505])$.

To explore the effects of deformation for the SS and PSS, in Fig.~\ref{Fig3},
the single-particle Nilsson levels $\Omega[\mathcal{N}n_{z}\Lambda]$ as a
function of deformation parameter $\beta$ is plotted ranging from $\beta=-0.4$
to $0.6$. The solid and dashed lines are respectively for the levels with
positive and negative parities. In the spherical case with $\beta=0$, distinct
shell structure emerges with the traditional magic numbers $2, 8, 20, 28, 50$,
and $82$ and seven pairs of spin doublets, i.e., $1p, 1d, 1f, 2p, 1g, 2d$, and
$1h$, are obtained. With the quadrupole deformations, as a result of levels
splitting, more pairs of spin doublets are obtained as a combination of
Nilsson levels with quantum
numbers~$\Omega=\Lambda-1/2~[\mathcal{N},n_{3},\Lambda]$ and
$\Omega=\Lambda+1/2~[\mathcal{N},n_{3},\Lambda]$. On closer inspection, in the
side of prolate deformations with $\beta$ ranging from $0$ to $0.6$, the
single-particle levels of spin doublets are approximately parallel, such as
the spin doublets $1d~(3/2[202], 5/2[202])$,  $1f~(5/2[303], 7/2[303])$, and
$2p~(1/2[301], 3/2[301])$. However, this relationship does not exist well in
the side of $\beta<0$ with oblate deformation.

\begin{figure}[t!]
  \includegraphics[width=0.45\textwidth]{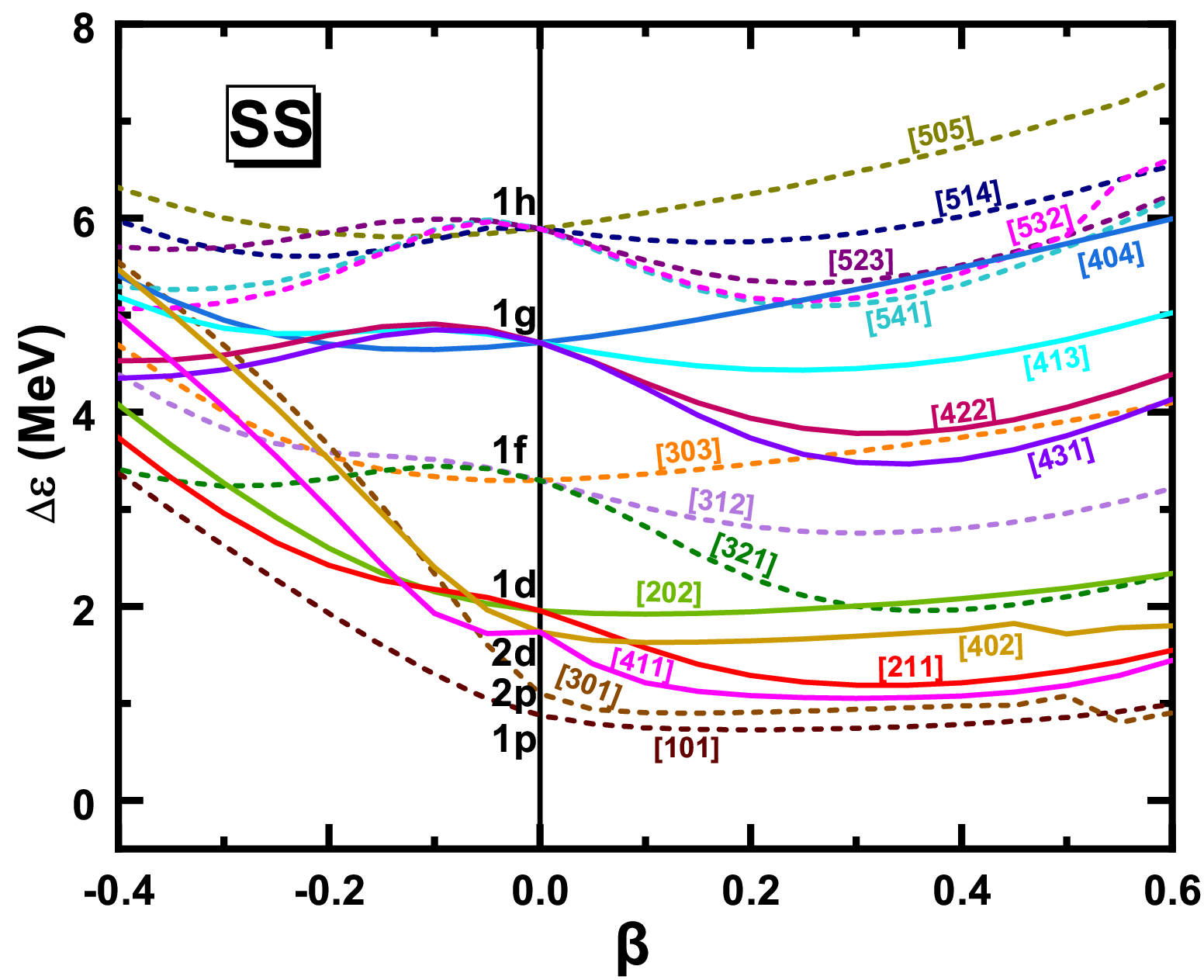}
\caption{Energy splittings $\Delta \varepsilon$ between the spin doublets
$\Lambda\pm 1/2~[\mathcal{N}n_{z}\Lambda]$ as a function of the deformation
parameter $\beta$. The solid and dashed lines indicate doublets with positive
and negative parities, respectively.}
  \label{Fig4}
\end{figure}

In order to better master the SS in the deformed nuclei, in Fig.~\ref{Fig4},
the energy splittings $\Delta
\varepsilon=\varepsilon_{\Omega_<}-\varepsilon_{\Omega_>}$ for the spin
doublets $\Lambda\pm1/2~[\mathcal{N},n_{3},\Lambda]$ are plotted as a function
of deformation $\beta$, where $\varepsilon_{\Omega_<}$ and
$\varepsilon_{\Omega_>}$ are the energies of the spin down and up states,
respectively. According to Fig.~\ref{Fig4}, the following points could be
revealed: (i) The energy splitting $\Delta \varepsilon$ between the spin
doublets for the bound states keep positive over the whole range of
deformation considered here. (ii) In the case of $\beta\geq 0$, the energy
splitting between spin doublet with the same principal quantum number~$n$ is
relatively larger for those owning larger orbital angular momentum $l$, i.e.,
$\Delta\varepsilon(1h)>\Delta\varepsilon(1g)>\Delta\varepsilon(1f)>\Delta\varepsilon(1d)>\Delta\varepsilon(1p)$
and $\Delta\varepsilon(2d)>\Delta\varepsilon(2p)$. (iii) For the spin doublets
with the same $n$ and $l$, the energy splitting with greater $\Lambda$ is
larger, e.g., for the $1h$ spin doublets, $\Delta
\varepsilon_{7/2[404]-9/2[404]}> \Delta \varepsilon_{5/2[413]-7/2[413]}>
\Delta \varepsilon_{3/2[422]-5/2[422]}
> \Delta \varepsilon_{1/2[431]-3/2[431]}$. (iv) The sensitivity of
energy splitting of the spin doublets on the deformation $\beta$ is much
higher in the oblate side than in the prolate side, especially for those with
lower angular momentum such as $1p$ and $2p$ spin doublets.

\begin{figure}[t!]
\includegraphics[width=0.45\textwidth]{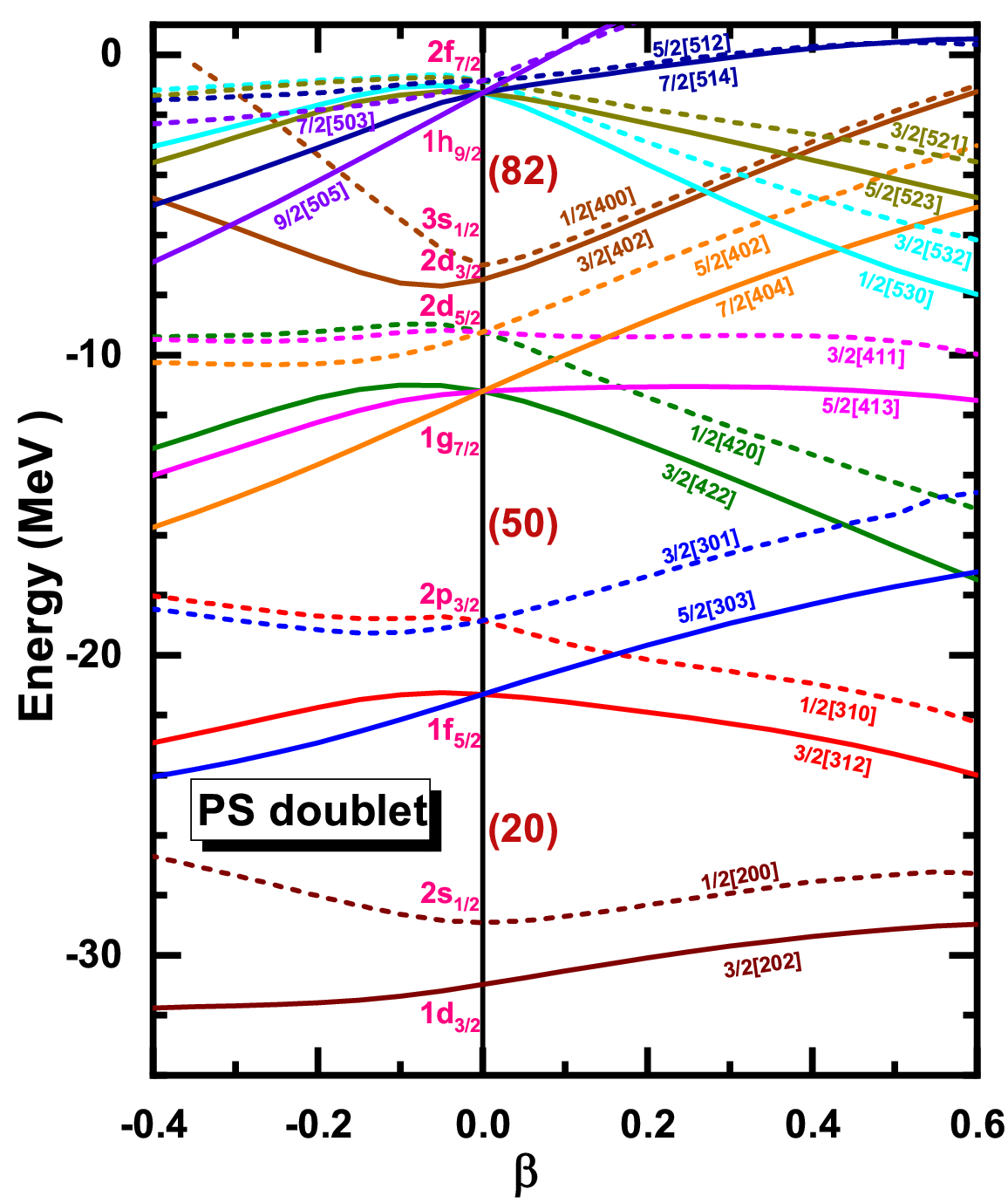}
\caption{Single-particle levels for the pseudospin doublets in $^{154}$Dy as a
function of deformation parameter $\beta$ in quadrupole-deformed Woods-Saxon
potentials. The solid and dashed lines with the same color denote one pseudospin doublet with the pseudospin $\tilde{s}=\pm 1/2$, respectively.}
\label{Fig5}
\end{figure}

In Fig.~\ref{Fig5}, the single-particle levels for all pseudospin doublets in
$^{154}$Dy are plotted as a function of deformation parameter $\beta$. The
solid and dashed lines with the same color denote one pair of PS doublet with
the pseudospin $\tilde{s}=\pm 1/2$, respectively. At spherical deformation,
five pairs of pseudospin doublets, i.e., $1\widetilde{p}~(2s_{1/2},
1d_{3/2})$, $1\widetilde{d}~(2p_{3/2}, 1f_{5/2})$, $1\widetilde{f}~(2d_{5/2},
1g_{7/2})$, $2\widetilde{p}~(3s_{1/2}, 2d_{5/2})$, and
$1\widetilde{g}~(2f_{7/2}, 1h_{9/2})$, are obtained. In the deformed case,
levels are split and more pairs of pseudospin doublets are obtained as a
combination of Nilsson levels with quantum
numbers~$\Omega=\Lambda+1/2~[\mathcal{N}n_{3}\Lambda]$ and
$\Omega=\Lambda+3/2~[\mathcal{N},n_{3},\Lambda+2]$, which correspond to the
pseudospin down state
$\Omega=\widetilde{\Lambda}-1/2~[\widetilde{\mathcal{N}n_{3}\Lambda]}$ and the
pseudospin up state
$\Omega=\widetilde{\Lambda}+1/2~[\widetilde{\mathcal{N}n_{3}\Lambda]}$. For
most of the bound pseudospin doublets, the pseudospin up state locates lower
than the pseudospin down state except those close to the continuum threshold
and those with a resonant state being a PS partner. At the deformation of
$\beta=0.1$, an energy level crossing between the doublets $7/2[503]$ and
$9/2[505]$ is observed, and the same phenomenon happens for the doublets
$5/2[512]$ and $7/2[514]$ around $\beta=0.5$. Such level crossings lead to the
change of the energy splittings between PS doublets, and in
Refs.~\cite{Bohr_PhysScr_1982,Lalazissis_PRC_1998,Sun_PRC_2019_034310}, the
same phenomenon has also been observed. Further studies has pointed out that
the reversed level structure is decided by the sign of the integration of the
pseudospin-orbit potential over $r$~\cite{Meng_PRC_1999}. This can also be
explained by the spin-orbit effects within the framework of supersymmetric
quantum mechanics as discussed in Refs.~\cite{Liang_PRC_2013,Shen_PRC_2013}.
Besides, almost for all the bound pseudospin doublets, the corresponding
single-particle levels approximately satisfy the parallel relationship in the
area of $\beta>0$, which is the same as spin doublets.

\begin{figure}[t!]
  \includegraphics[width=0.45\textwidth]{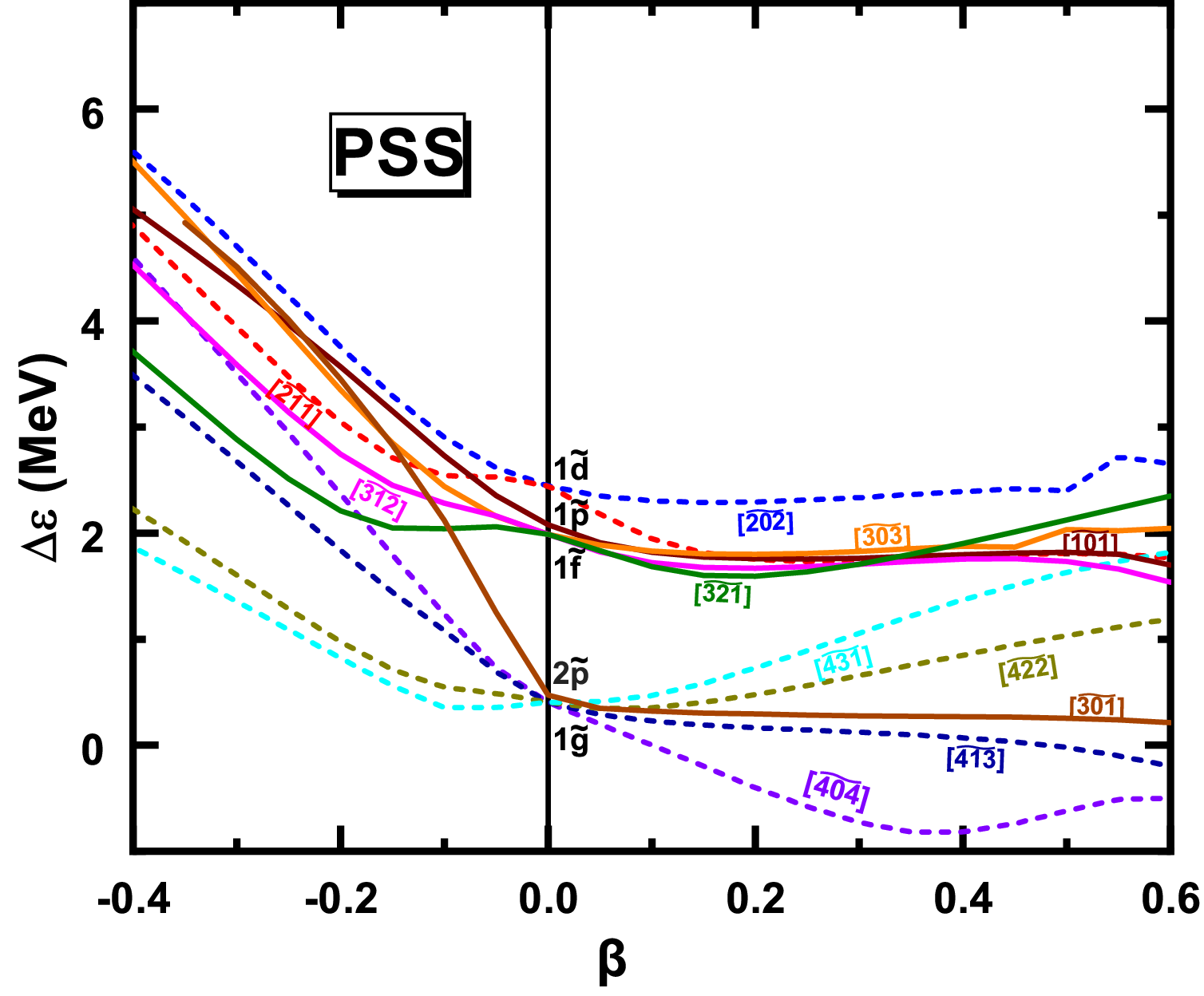}
\caption{The same as Fig.~\ref{Fig4}, but for the pseudospin doublets
$\widetilde{\Lambda}\pm 1/2~[\widetilde{\mathcal{N}n_{z}\Lambda}]$.}
  \label{Fig6}
\end{figure}
\begin{figure*}[t!]
  \includegraphics[width=0.7\textwidth]{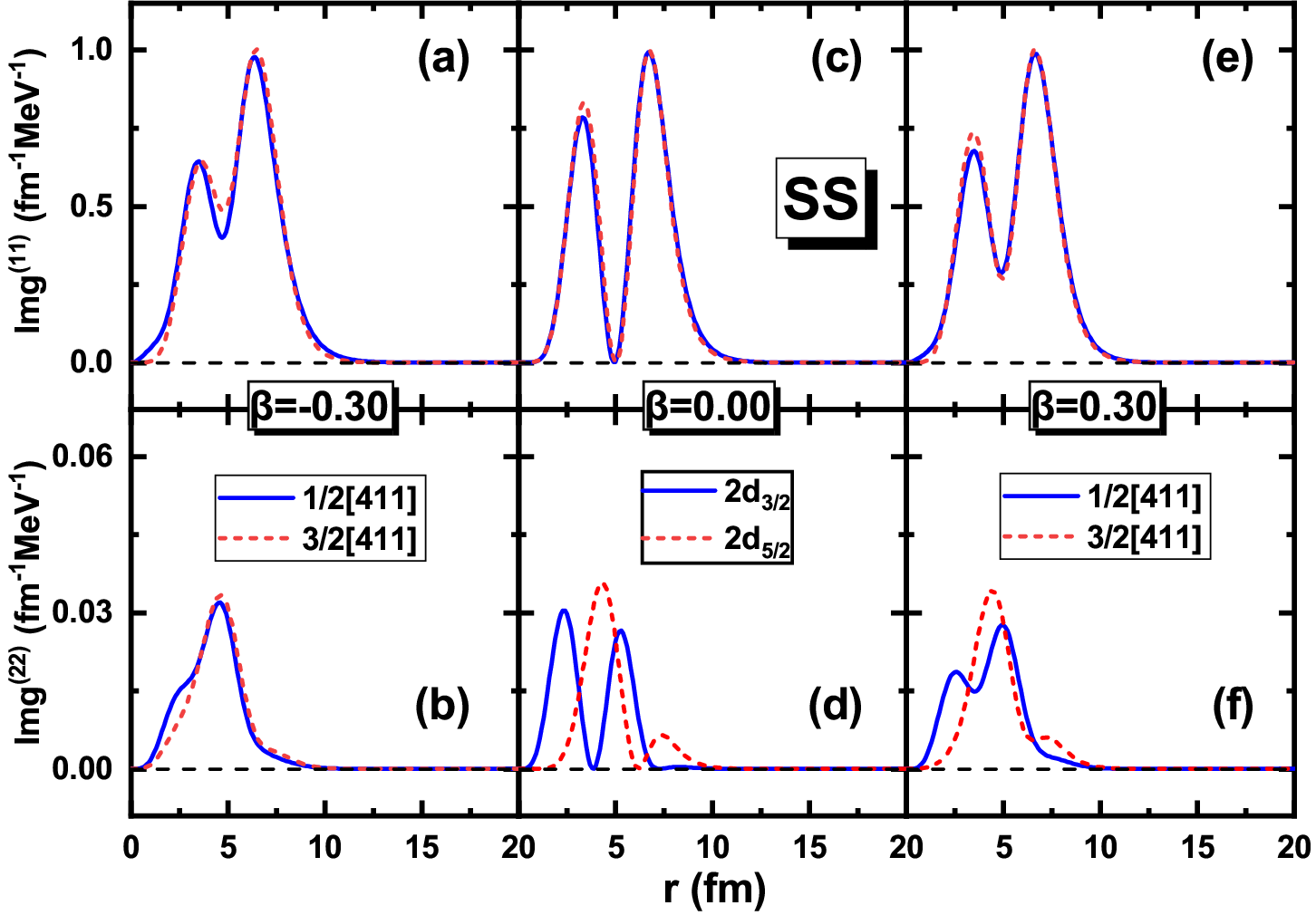}
  \caption{Density distributions in the coordinate space, Im${\mathcal{G}}_{\kappa}^{(11)}$ and Im${\mathcal{G}}_{\kappa}^{(22)}$,
 for the $2d~(1/2[411], 3/2[411])$ spin doublets with different deformation parameters $\beta=-0.30$~(a,b), $0.00$~(c,d), and $0.30$~(e,f).}
  \label{Fig7}
\end{figure*}
\begin{figure*}[ht!]
  \includegraphics[width=0.7\textwidth]{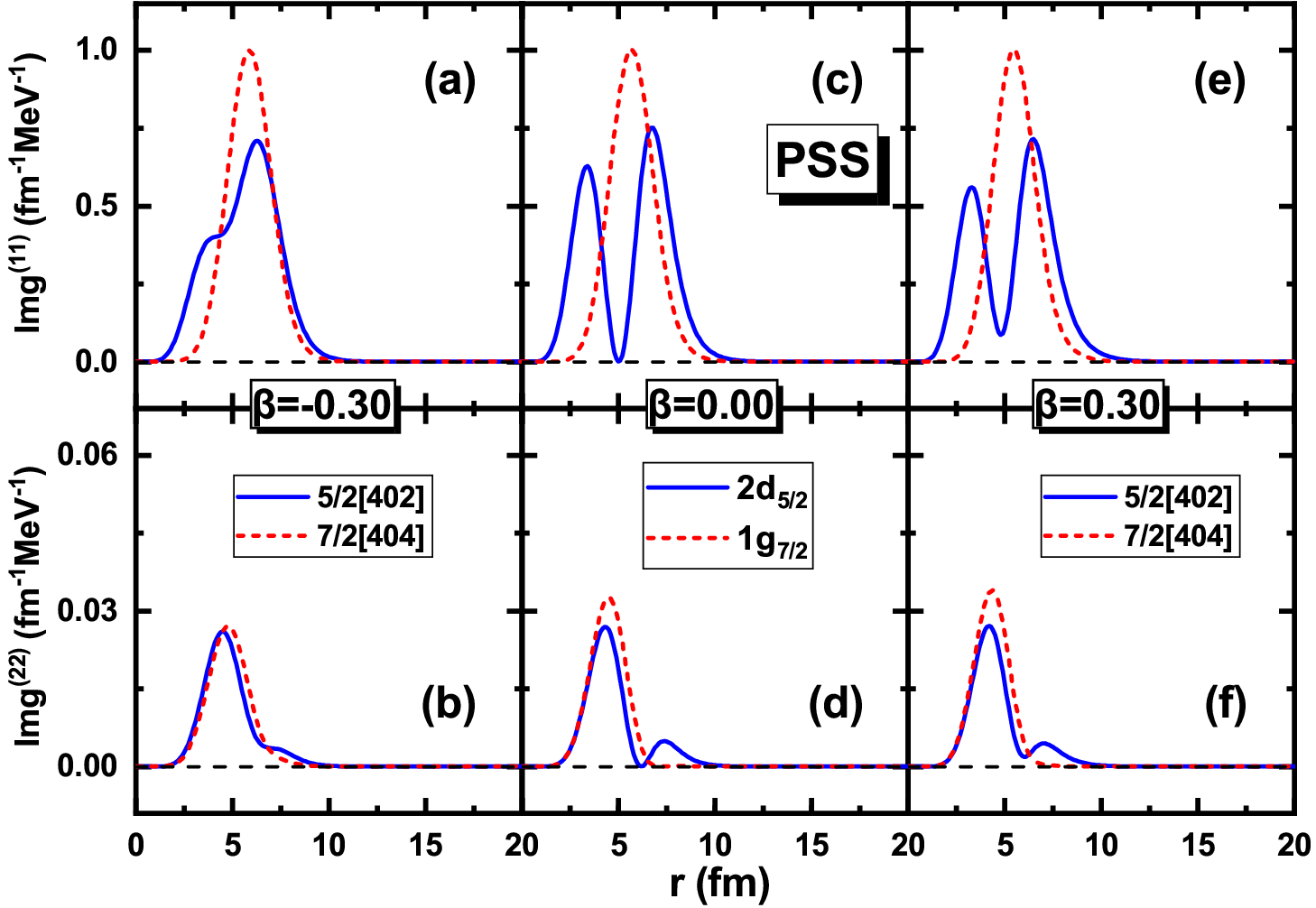}
  \caption{The same as Fig.~\ref{Fig7}, but for the pseudospin doublet $1\widetilde{f}~(5/2[402], 7/2[404])$.}
  \label{Fig8}
\end{figure*}

Similarly as the spin doublets, in Fig.~\ref{Fig6}, the energy splittings
$\Delta\varepsilon=\varepsilon_{\widetilde{\Omega}_<}-\varepsilon_{\widetilde{\Omega}_>}$
between the pseudospin doublets are plotted as a function of the deformation
parameter $\beta$, where $\varepsilon_{\widetilde{\Omega}_<}$ and
$\varepsilon_{\widetilde{\Omega}_>}$ are energies of the pseudospin down and
up doublets, respectively. For convenience, we use the pseudo quantum numbers
$\widetilde{[\mathcal{N} n_{z} \Lambda]}$ to denote the PS doublet
$\widetilde{\Lambda}\mp
1/2~[\widetilde{\mathcal{N}}=\mathcal{N}-1,\widetilde{n}_{3}={n}_{3},\widetilde{\Lambda}=\Lambda+1]$
with a combination of $\Lambda+1/2~[\mathcal{N},n_{3},\Lambda]$ and
$\Lambda+3/2~[\mathcal{N},n_{3},\Lambda+2]$. For example, the PS doublet
$(1/2[200], 3/2[202])$ can be denoted as $[\widetilde{101}]$. From
Figs.~\ref{Fig5} and \ref{Fig6}, the following points can be revealed for the
PSS: (i) The energy difference between the pseudospin up and down states
always remains positive except for the pseudospin doublets $\widetilde{[413]}$
and $\widetilde{[404]}$, the partners of which have crossed the continuum
threshold. (ii) The energy splitting is much smaller for the pseudospin
doublets close to continuum threshold, revealing the good pseudospin symmetry,
which has also been observed in Refs.~\cite{Guo_PRL_2014,Lalazissis_PRC_1998}.
(iii) In the side of prolate deformation with $\beta>0$, the energy splitting
between PS parters keeps almost constant for most of the pseudospin doublets.

For a pair of (pseudo)spin doublet, the good (P)SS results in not only a
(quasi)degeneracy in their energies, but also great similarities in their
Dirac wave functions. For the spin doublets, the upper components of the Dirac
spinor $G(r)$ behave similarly, and for the pseudospin doublets, the lower
components of the Dirac spinor $F(r)$ are similar. In the PSS limit, i.e.,
when the attractive scalar and repulsive vector potentials have the same
magnitude but opposite sign, identical density distributions of the lower
components have been proved for the PS partners~\cite{PLB2023TTSun}. With the
Green's function method, the density distributions in the coordinate space can
be used to study the similarities of Dirac spinors and in the deformed nuclei,
the radial density distributions corresponding to the upper and lower
components of the Dirac wave functions for the single-particle state with
$\Omega^\pi$ at the eigenvalues of $\varepsilon$ could be calculated by
\begin{eqnarray}
&&\rho^{(g)}_{\Omega}(r,\varepsilon) = -\frac{1}{4\pi r^2}\sum\limits_{\kappa}{\rm
Im}{\mathcal{G}}_{\Omega\kappa\kappa}^{(11)}(r,r;\varepsilon),\\
&&\rho^{(f)}_{\Omega}(r,\varepsilon) = -\frac{1}{4\pi r^2}\sum\limits_{\kappa}{\rm
Im}{\mathcal{G}}_{\Omega\kappa\kappa}^{(22)}(r,r;\varepsilon),
\end{eqnarray}
where different spherical partial waves are coupled together.

In Fig.~\ref{Fig7}, taking the spin doublets $2d(1/2[411], 3/2[411])$ as
examples, the density distributions $\rho_{\Omega}^{(g)}(r,\varepsilon)$ and
$\rho_{\Omega}^{(f)}(r,\varepsilon)$ are plotted at different deformations
$\beta=-0.3$, $0.0$, $0.3$. In the spherical case, great similarities are
observed in ${\rm Im}{\mathcal{G}}_{\kappa}^{(11)}$ corresponding to the upper
component of Dirac wave functions while a big difference exists in the lower
component ${\rm Im}{\mathcal{G}}_{\kappa}^{(22)}$ between the spin partners.
All those together with a small energy splitting support the approximate spin
symmetry. When the mean field potential deviates from the spherical shape,
good consistency in ${\rm Im}{\mathcal{G}}_{\kappa}^{(11)}$ between the spin
doublets still keeps at prolate deformation of $\beta = 0.30$ while some more
differences happen at oblate deformation of $\beta=-0.3$. As seen in
Fig.~\ref{Fig4}, the energy splitting between the spin doublets $2d (1/2[411],
3/2[411])$ is small in the side of prolate deformation and it keeps stable
with the increase of $\beta$ from $0$ to $0.6$, while in the oblate side,
great differences are revealed at various deformations for the SS doublets
especially for those with low angular momentum.

In Fig~\ref{Fig8}, the density distributions of the PSS doublets
$1\widetilde{f} (5/2[402], 7/2[404])$ at different deformations $\beta=-0.3,
0.0, 0.3$. In the spherical case, some similarities are revealed in the
density distributions Im${\mathcal{G}}_{\kappa}^{(22)}(r,\varepsilon)$ of the
PSS doublets while one node difference, i.e., $n(2d_{5/2})=n(1g_{7/2})+1$, is
observed in the density distributions
Im${\mathcal{G}}_{\kappa}^{(11)}(r,\varepsilon)$ corresponding to the upper
component of the Dirac wave functions. In the deformed case, the similarities
in the lower component of the Dirac wave function keep well while the node
relationship in the upper component
Im${\mathcal{G}}_{\kappa}^{(11)}(r,\varepsilon)$ is no longer satisfied.

\section{Summary}
\label{Sec:Summary}

In this work, Green's function method has been
applied to examine the possible spin and pseudospin symmetries in deformed
nuclei for the first time, which provides a novel way to exactly determine the
single-particle energies and also properly describe the density
distributions.

Firstly, taking the axially deformed nucleus $^{154}$Dy as an example, the
density of states are calculated with the Green's functions method by solving
the coupled-channel Dirac equation with quadrupole-deformed Woods-Saxon
potentials. By searching for the extremes of DoSs, the single-particle
energies for $^{154}$Dy are exactly obtained, based on which the spin doublets
$\Lambda\pm 1/2[\mathcal{N},n_z, \Lambda]$ and pseudospin doublets
$\widetilde{\Lambda}\pm 1/2[\widetilde{\mathcal{N},n_z, \Lambda}]$ are
determined. Different behaviors are displayed for the spin and pseudospin
doublets. SS for doublets with the same $\Lambda$ deteriorates with increasing
$l$ and for those with the same $n$ and $l$, it also becomes worse for larger
$\Lambda$. Differently, PSS for doublets with the same $\widetilde{\Lambda}$
is better maintained with increasing $\widetilde{l}$ and for those with the
same $\widetilde{n}$ and $\widetilde{l}$, it sometimes becomes better for
larger $\widetilde{\Lambda}$. Besides, great threshold effect is exhibited for
PSS. One common point for the spin and pseudospin doublets is the inversions
of energy levels between the partners observed in the area of continuum.

Secondly, to explore the effects of deformation for the SS and PSS, the
single-particle Nilsson levels $\Omega[\mathcal{N},n_z,\Lambda]$ are plotted
as a function of deformation ranging from $\beta=-0.4$ to $0.6$. By studying
the energy splittings between the partners, the conservation and breaking of
SS and PSS are examined at different deformations. In the side of prolate
deformations with $\beta > 0$, the Nilsson levels for the spin and pseudospin
doublets are almost parallel and the energy splittings are stable against
different deformations. However, the energy splitting is much sensitive to the
deformation $\beta$ in the oblate side with $\beta < 0$. For the spin
doublets, the energy splitting for the bound states keep positive over the
whole range of deformation. The same conclusion is obtained for pseudospin
doublets except those close to the continuum threshold. Besides, good spin
symmetry may appear for partners with smaller angular momentum such as $1p$
partners while good pseudospin symmetry appears in the states which locate
close to the continuum threshold.

Finally, the density distributions in the coordinate space are plotted for the
spin and pseudospin doublets at different deformations $\beta=-0.3, 0, 0.3$.
For the spin doublets, great similarities are observed for the
$\rho_\Omega^{(g)}(r,\varepsilon)$ which is related with the upper component
of the Dirac wave functions while great similarities are observed for the
density distribution $\rho_\Omega^{(f)}(r,\varepsilon)$ which related with the
lower component of the Dirac wave function for the pseudospin doublets.
Besides, these similarities can be maintained well at different deformations.

\begin{acknowledgments}
This work was partly supported by the National Natural Science Foundation of
China (No. U2032141), the Open Project of Guangxi Key Laboratory of Nuclear
Physics and Nuclear Technology (No. NLK2022-02), the Central Government
Guidance Funds for Local Scientific and Technological Development, China (No.
Guike ZY22096024), the Foundation of Fundamental Research for Young Teachers
of Zhengzhou University (JC202041041), and the Physics Research and
Development Program of Zhengzhou University (32410217). The theoretical
calculation was supported by the nuclear data storage system in Zhengzhou
University.
\end{acknowledgments}


\begin{thebibliography}{71}%
\makeatletter
\providecommand \@ifxundefined [1]{%
 \@ifx{#1\undefined}
}%
\providecommand \@ifnum [1]{%
 \ifnum #1\expandafter \@firstoftwo
 \else \expandafter \@secondoftwo
 \fi
}%
\providecommand \@ifx [1]{%
 \ifx #1\expandafter \@firstoftwo
 \else \expandafter \@secondoftwo
 \fi
}%
\providecommand \natexlab [1]{#1}%
\providecommand \enquote  [1]{``#1''}%
\providecommand \bibnamefont  [1]{#1}%
\providecommand \bibfnamefont [1]{#1}%
\providecommand \citenamefont [1]{#1}%
\providecommand \href@noop [0]{\@secondoftwo}%
\providecommand \href [0]{\begingroup \@sanitize@url \@href}%
\providecommand \@href[1]{\@@startlink{#1}\@@href}%
\providecommand \@@href[1]{\endgroup#1\@@endlink}%
\providecommand \@sanitize@url [0]{\catcode `\\12\catcode `\$12\catcode
  `\&12\catcode `\#12\catcode `\^12\catcode `\_12\catcode `\%12\relax}%
\providecommand \@@startlink[1]{}%
\providecommand \@@endlink[0]{}%
\providecommand \url  [0]{\begingroup\@sanitize@url \@url }%
\providecommand \@url [1]{\endgroup\@href {#1}{\urlprefix }}%
\providecommand \urlprefix  [0]{URL }%
\providecommand \Eprint [0]{\href }%
\providecommand \doibase [0]{http://dx.doi.org/}%
\providecommand \selectlanguage [0]{\@gobble}%
\providecommand \bibinfo  [0]{\@secondoftwo}%
\providecommand \bibfield  [0]{\@secondoftwo}%
\providecommand \translation [1]{[#1]}%
\providecommand \BibitemOpen [0]{}%
\providecommand \bibitemStop [0]{}%
\providecommand \bibitemNoStop [0]{.\EOS\space}%
\providecommand \EOS [0]{\spacefactor3000\relax}%
\providecommand \BibitemShut  [1]{\csname bibitem#1\endcsname}%
\let\auto@bib@innerbib\@empty
\bibitem [{\citenamefont {Ginocchio}(2005)}]{Ginocchio_PhysRep_2005}%
  \BibitemOpen
  \bibfield  {author} {\bibinfo {author} {\bibfnamefont {J.~N.}\ \bibnamefont
  {Ginocchio}},\ }\href {\doibase
  https://doi.org/10.1016/j.physrep.2005.04.003} {\bibfield  {journal}
  {\bibinfo  {journal} {Phys. Rep.}\ }\textbf {\bibinfo {volume} {414}},\
  \bibinfo {pages} {165} (\bibinfo {year} {2005})}\BibitemShut {NoStop}%
\bibitem [{\citenamefont {Liang}\ \emph {et~al.}(2015)\citenamefont {Liang},
  \citenamefont {Meng},\ and\ \citenamefont {Zhou}}]{Liang_PhysRep_2015}%
  \BibitemOpen
  \bibfield  {author} {\bibinfo {author} {\bibfnamefont {H.}~\bibnamefont
  {Liang}}, \bibinfo {author} {\bibfnamefont {J.}~\bibnamefont {Meng}}, \ and\
  \bibinfo {author} {\bibfnamefont {S.-G.}\ \bibnamefont {Zhou}},\ }\href
  {\doibase https://doi.org/10.1016/j.physrep.2014.12.005} {\bibfield
  {journal} {\bibinfo  {journal} {Phys. Rep.}\ }\textbf {\bibinfo {volume}
  {570}},\ \bibinfo {pages} {1} (\bibinfo {year} {2015})}\BibitemShut {NoStop}%
\bibitem [{\citenamefont {Leviatan}\ and\ \citenamefont
  {Ginocchio}(2001)}]{Leviatan_PLB_2001}%
  \BibitemOpen
  \bibfield  {author} {\bibinfo {author} {\bibfnamefont {A.}~\bibnamefont
  {Leviatan}}\ and\ \bibinfo {author} {\bibfnamefont {J.~N.}\ \bibnamefont
  {Ginocchio}},\ }\href {\doibase
  https://doi.org/10.1016/S0370-2693(01)01039-5} {\bibfield  {journal}
  {\bibinfo  {journal} {Phys. Lett. B}\ }\textbf {\bibinfo {volume} {518}},\
  \bibinfo {pages} {214} (\bibinfo {year} {2001})}\BibitemShut {NoStop}%
\bibitem [{\citenamefont {Shen}\ \emph {et~al.}(2018)\citenamefont {Shen},
  \citenamefont {Liang}, \citenamefont {Meng}, \citenamefont {Ring},\ and\
  \citenamefont {Zhang}}]{Shen_PLB_2018}%
  \BibitemOpen
  \bibfield  {author} {\bibinfo {author} {\bibfnamefont {S.}~\bibnamefont
  {Shen}}, \bibinfo {author} {\bibfnamefont {H.}~\bibnamefont {Liang}},
  \bibinfo {author} {\bibfnamefont {J.}~\bibnamefont {Meng}}, \bibinfo {author}
  {\bibfnamefont {P.}~\bibnamefont {Ring}}, \ and\ \bibinfo {author}
  {\bibfnamefont {S.}~\bibnamefont {Zhang}},\ }\href {\doibase
  https://doi.org/10.1016/j.physletb.2018.03.080} {\bibfield  {journal}
  {\bibinfo  {journal} {Phys. Lett. B}\ }\textbf {\bibinfo {volume} {781}},\
  \bibinfo {pages} {227} (\bibinfo {year} {2018})}\BibitemShut {NoStop}%
\bibitem [{\citenamefont {Haxel}\ \emph {et~al.}(1949)\citenamefont {Haxel},
  \citenamefont {Jensen},\ and\ \citenamefont {Suess}}]{Otto_PhysRev_1949}%
  \BibitemOpen
  \bibfield  {author} {\bibinfo {author} {\bibfnamefont {O.}~\bibnamefont
  {Haxel}}, \bibinfo {author} {\bibfnamefont {J.~H.~D.}\ \bibnamefont
  {Jensen}}, \ and\ \bibinfo {author} {\bibfnamefont {H.~E.}\ \bibnamefont
  {Suess}},\ }\href {\doibase 10.1103/PhysRev.75.1766.2} {\bibfield  {journal}
  {\bibinfo  {journal} {Phys. Rev.}\ }\textbf {\bibinfo {volume} {75}},\
  \bibinfo {pages} {1766} (\bibinfo {year} {1949})}\BibitemShut {NoStop}%
\bibitem [{\citenamefont {Mayer}(1949)}]{Mayer_PhysRev_1949}%
  \BibitemOpen
  \bibfield  {author} {\bibinfo {author} {\bibfnamefont {M.~G.}\ \bibnamefont
  {Mayer}},\ }\href {\doibase 10.1103/PhysRev.75.1969} {\bibfield  {journal}
  {\bibinfo  {journal} {Phys. Rev.}\ }\textbf {\bibinfo {volume} {75}},\
  \bibinfo {pages} {1969} (\bibinfo {year} {1949})}\BibitemShut {NoStop}%
\bibitem [{\citenamefont {Hecht}\ and\ \citenamefont
  {Adler}(1969)}]{Hecht_NPA_1969}%
  \BibitemOpen
  \bibfield  {author} {\bibinfo {author} {\bibfnamefont {K.}~\bibnamefont
  {Hecht}}\ and\ \bibinfo {author} {\bibfnamefont {A.}~\bibnamefont {Adler}},\
  }\href {\doibase https://doi.org/10.1016/0375-9474(69)90077-3} {\bibfield
  {journal} {\bibinfo  {journal} {Nucl. Phys. A}\ }\textbf {\bibinfo {volume}
  {137}},\ \bibinfo {pages} {129} (\bibinfo {year} {1969})}\BibitemShut
  {NoStop}%
\bibitem [{\citenamefont {Arima}\ \emph {et~al.}(1969)\citenamefont {Arima},
  \citenamefont {Harvey},\ and\ \citenamefont {Shimizu}}]{Arima_PLB_1969}%
  \BibitemOpen
  \bibfield  {author} {\bibinfo {author} {\bibfnamefont {A.}~\bibnamefont
  {Arima}}, \bibinfo {author} {\bibfnamefont {M.}~\bibnamefont {Harvey}}, \
  and\ \bibinfo {author} {\bibfnamefont {K.}~\bibnamefont {Shimizu}},\ }\href
  {\doibase https://doi.org/10.1016/0370-2693(69)90443-2} {\bibfield  {journal}
  {\bibinfo  {journal} {Phys. Lett. B}\ }\textbf {\bibinfo {volume} {30}},\
  \bibinfo {pages} {517} (\bibinfo {year} {1969})}\BibitemShut {NoStop}%
\bibitem [{\citenamefont {Bohr}\ \emph {et~al.}(1982)\citenamefont {Bohr},
  \citenamefont {Hamamoto},\ and\ \citenamefont
  {Mottelson}}]{Bohr_PhysScr_1982}%
  \BibitemOpen
  \bibfield  {author} {\bibinfo {author} {\bibfnamefont {A.}~\bibnamefont
  {Bohr}}, \bibinfo {author} {\bibfnamefont {I.}~\bibnamefont {Hamamoto}}, \
  and\ \bibinfo {author} {\bibfnamefont {B.~R.}\ \bibnamefont {Mottelson}},\
  }\href {\doibase 10.1088/0031-8949/26/4/003} {\bibfield  {journal} {\bibinfo
  {journal} {Phys. Scr.}\ }\textbf {\bibinfo {volume} {26}},\ \bibinfo {pages}
  {267} (\bibinfo {year} {1982})}\BibitemShut {NoStop}%
\bibitem [{\citenamefont {Dudek}\ \emph {et~al.}(1987)\citenamefont {Dudek},
  \citenamefont {Nazarewicz}, \citenamefont {Szymanski},\ and\ \citenamefont
  {Leander}}]{Dudek_PRL_1987}%
  \BibitemOpen
  \bibfield  {author} {\bibinfo {author} {\bibfnamefont {J.}~\bibnamefont
  {Dudek}}, \bibinfo {author} {\bibfnamefont {W.}~\bibnamefont {Nazarewicz}},
  \bibinfo {author} {\bibfnamefont {Z.}~\bibnamefont {Szymanski}}, \ and\
  \bibinfo {author} {\bibfnamefont {G.~A.}\ \bibnamefont {Leander}},\ }\href
  {\doibase 10.1103/PhysRevLett.59.1405} {\bibfield  {journal} {\bibinfo
  {journal} {Phys. Rev. Lett.}\ }\textbf {\bibinfo {volume} {59}},\ \bibinfo
  {pages} {1405} (\bibinfo {year} {1987})}\BibitemShut {NoStop}%
\bibitem [{\citenamefont {Nazarewicz}\ \emph {et~al.}(1990)\citenamefont
  {Nazarewicz}, \citenamefont {Twin}, \citenamefont {Fallon},\ and\
  \citenamefont {Garrett}}]{Nazarewicz_PRL_1990}%
  \BibitemOpen
  \bibfield  {author} {\bibinfo {author} {\bibfnamefont {W.}~\bibnamefont
  {Nazarewicz}}, \bibinfo {author} {\bibfnamefont {P.~J.}\ \bibnamefont
  {Twin}}, \bibinfo {author} {\bibfnamefont {P.}~\bibnamefont {Fallon}}, \ and\
  \bibinfo {author} {\bibfnamefont {J.~D.}\ \bibnamefont {Garrett}},\ }\href
  {\doibase 10.1103/PhysRevLett.64.1654} {\bibfield  {journal} {\bibinfo
  {journal} {Phys. Rev. Lett.}\ }\textbf {\bibinfo {volume} {64}},\ \bibinfo
  {pages} {1654} (\bibinfo {year} {1990})}\BibitemShut {NoStop}%
\bibitem [{\citenamefont {Byrski}\ \emph {et~al.}(1990)\citenamefont {Byrski},
  \citenamefont {Beck}, \citenamefont {Curien}, \citenamefont {Schuck},
  \citenamefont {Fallon}, \citenamefont {Alderson}, \citenamefont {Ali},
  \citenamefont {Bentley}, \citenamefont {Bruce}, \citenamefont {Forsyth},
  \citenamefont {Howe}, \citenamefont {Roberts}, \citenamefont
  {Sharpey-Schafer}, \citenamefont {Smith},\ and\ \citenamefont
  {Twin}}]{Byrski_PRL_1990}%
  \BibitemOpen
  \bibfield  {author} {\bibinfo {author} {\bibfnamefont {T.}~\bibnamefont
  {Byrski}}, \bibinfo {author} {\bibfnamefont {F.~A.}\ \bibnamefont {Beck}},
  \bibinfo {author} {\bibfnamefont {D.}~\bibnamefont {Curien}}, \bibinfo
  {author} {\bibfnamefont {C.}~\bibnamefont {Schuck}}, \bibinfo {author}
  {\bibfnamefont {P.}~\bibnamefont {Fallon}}, \bibinfo {author} {\bibfnamefont
  {A.}~\bibnamefont {Alderson}}, \bibinfo {author} {\bibfnamefont
  {I.}~\bibnamefont {Ali}}, \bibinfo {author} {\bibfnamefont {M.~A.}\
  \bibnamefont {Bentley}}, \bibinfo {author} {\bibfnamefont {A.~M.}\
  \bibnamefont {Bruce}}, \bibinfo {author} {\bibfnamefont {P.~D.}\ \bibnamefont
  {Forsyth}}, \bibinfo {author} {\bibfnamefont {D.}~\bibnamefont {Howe}},
  \bibinfo {author} {\bibfnamefont {J.~W.}\ \bibnamefont {Roberts}}, \bibinfo
  {author} {\bibfnamefont {J.~F.}\ \bibnamefont {Sharpey-Schafer}}, \bibinfo
  {author} {\bibfnamefont {G.}~\bibnamefont {Smith}}, \ and\ \bibinfo {author}
  {\bibfnamefont {P.~J.}\ \bibnamefont {Twin}},\ }\href {\doibase
  10.1103/PhysRevLett.64.1650} {\bibfield  {journal} {\bibinfo  {journal}
  {Phys. Rev. Lett.}\ }\textbf {\bibinfo {volume} {64}},\ \bibinfo {pages}
  {1650} (\bibinfo {year} {1990})}\BibitemShut {NoStop}%
\bibitem [{\citenamefont {Castanos}\ \emph {et~al.}(1992)\citenamefont
  {Castanos}, \citenamefont {Moshinsky},\ and\ \citenamefont
  {Quesne}}]{Castanos_PLB_1992}%
  \BibitemOpen
  \bibfield  {author} {\bibinfo {author} {\bibfnamefont {O.}~\bibnamefont
  {Castanos}}, \bibinfo {author} {\bibfnamefont {M.}~\bibnamefont {Moshinsky}},
  \ and\ \bibinfo {author} {\bibfnamefont {C.}~\bibnamefont {Quesne}},\ }\href
  {\doibase https://doi.org/10.1016/0370-2693(92)90741-L} {\bibfield  {journal}
  {\bibinfo  {journal} {Phys. Lett. B}\ }\textbf {\bibinfo {volume} {277}},\
  \bibinfo {pages} {238} (\bibinfo {year} {1992})}\BibitemShut {NoStop}%
\bibitem [{\citenamefont {Bahri}\ \emph {et~al.}(1992)\citenamefont {Bahri},
  \citenamefont {Draayer},\ and\ \citenamefont {Moszkowski}}]{Bahri_PRL_1992}%
  \BibitemOpen
  \bibfield  {author} {\bibinfo {author} {\bibfnamefont {C.}~\bibnamefont
  {Bahri}}, \bibinfo {author} {\bibfnamefont {J.~P.}\ \bibnamefont {Draayer}},
  \ and\ \bibinfo {author} {\bibfnamefont {S.~A.}\ \bibnamefont {Moszkowski}},\
  }\href {\doibase 10.1103/PhysRevLett.68.2133} {\bibfield  {journal} {\bibinfo
   {journal} {Phys. Rev. Lett.}\ }\textbf {\bibinfo {volume} {68}},\ \bibinfo
  {pages} {2133} (\bibinfo {year} {1992})}\BibitemShut {NoStop}%
\bibitem [{\citenamefont {Blokhin}\ \emph {et~al.}(1995)\citenamefont
  {Blokhin}, \citenamefont {Bahri},\ and\ \citenamefont
  {Draayer}}]{Blokhin_PRL_1995}%
  \BibitemOpen
  \bibfield  {author} {\bibinfo {author} {\bibfnamefont {A.~L.}\ \bibnamefont
  {Blokhin}}, \bibinfo {author} {\bibfnamefont {C.}~\bibnamefont {Bahri}}, \
  and\ \bibinfo {author} {\bibfnamefont {J.~P.}\ \bibnamefont {Draayer}},\
  }\href {\doibase 10.1103/PhysRevLett.74.4149} {\bibfield  {journal} {\bibinfo
   {journal} {Phys. Rev. Lett.}\ }\textbf {\bibinfo {volume} {74}},\ \bibinfo
  {pages} {4149} (\bibinfo {year} {1995})}\BibitemShut {NoStop}%
\bibitem [{\citenamefont {Ginocchio}(1997)}]{Ginocchio_PRL_1997}%
  \BibitemOpen
  \bibfield  {author} {\bibinfo {author} {\bibfnamefont {J.~N.}\ \bibnamefont
  {Ginocchio}},\ }\href {\doibase 10.1103/PhysRevLett.78.436} {\bibfield
  {journal} {\bibinfo  {journal} {Phys. Rev. Lett.}\ }\textbf {\bibinfo
  {volume} {78}},\ \bibinfo {pages} {436} (\bibinfo {year} {1997})}\BibitemShut
  {NoStop}%
\bibitem [{\citenamefont {Ginocchio}\ and\ \citenamefont
  {Madland}(1998)}]{Ginocchio_PRC_1998}%
  \BibitemOpen
  \bibfield  {author} {\bibinfo {author} {\bibfnamefont {J.~N.}\ \bibnamefont
  {Ginocchio}}\ and\ \bibinfo {author} {\bibfnamefont {D.~G.}\ \bibnamefont
  {Madland}},\ }\href {\doibase 10.1103/PhysRevC.57.1167} {\bibfield  {journal}
  {\bibinfo  {journal} {Phys. Rev. C}\ }\textbf {\bibinfo {volume} {57}},\
  \bibinfo {pages} {1167} (\bibinfo {year} {1998})}\BibitemShut {NoStop}%
\bibitem [{\citenamefont {Meng}\ \emph {et~al.}(1998)\citenamefont {Meng},
  \citenamefont {Sugawara-Tanabe}, \citenamefont {Yamaji}, \citenamefont
  {Ring},\ and\ \citenamefont {Arima}}]{Meng_PRC_1998}%
  \BibitemOpen
  \bibfield  {author} {\bibinfo {author} {\bibfnamefont {J.}~\bibnamefont
  {Meng}}, \bibinfo {author} {\bibfnamefont {K.}~\bibnamefont
  {Sugawara-Tanabe}}, \bibinfo {author} {\bibfnamefont {S.}~\bibnamefont
  {Yamaji}}, \bibinfo {author} {\bibfnamefont {P.}~\bibnamefont {Ring}}, \ and\
  \bibinfo {author} {\bibfnamefont {A.}~\bibnamefont {Arima}},\ }\href
  {\doibase 10.1103/PhysRevC.58.R628} {\bibfield  {journal} {\bibinfo
  {journal} {Phys. Rev. C}\ }\textbf {\bibinfo {volume} {58}},\ \bibinfo
  {pages} {R628} (\bibinfo {year} {1998})}\BibitemShut {NoStop}%
\bibitem [{\citenamefont {Meng}\ \emph {et~al.}(1999)\citenamefont {Meng},
  \citenamefont {Sugawara-Tanabe}, \citenamefont {Yamaji},\ and\ \citenamefont
  {Arima}}]{Meng_PRC_1999}%
  \BibitemOpen
  \bibfield  {author} {\bibinfo {author} {\bibfnamefont {J.}~\bibnamefont
  {Meng}}, \bibinfo {author} {\bibfnamefont {K.}~\bibnamefont
  {Sugawara-Tanabe}}, \bibinfo {author} {\bibfnamefont {S.}~\bibnamefont
  {Yamaji}}, \ and\ \bibinfo {author} {\bibfnamefont {A.}~\bibnamefont
  {Arima}},\ }\href {\doibase 10.1103/PhysRevC.59.154} {\bibfield  {journal}
  {\bibinfo  {journal} {Phys. Rev. C}\ }\textbf {\bibinfo {volume} {59}},\
  \bibinfo {pages} {154} (\bibinfo {year} {1999})}\BibitemShut {NoStop}%
\bibitem [{\citenamefont {Song}\ \emph {et~al.}(2009)\citenamefont {Song},
  \citenamefont {Yao},\ and\ \citenamefont {Meng}}]{Song_CPL_2009}%
  \BibitemOpen
  \bibfield  {author} {\bibinfo {author} {\bibfnamefont {C.-Y.}\ \bibnamefont
  {Song}}, \bibinfo {author} {\bibfnamefont {J.-M.}\ \bibnamefont {Yao}}, \
  and\ \bibinfo {author} {\bibfnamefont {J.}~\bibnamefont {Meng}},\ }\href
  {\doibase 10.1088/0256-307x/26/12/122102} {\bibfield  {journal} {\bibinfo
  {journal} {Chin. Phys. Lett.}\ }\textbf {\bibinfo {volume} {26}},\ \bibinfo
  {pages} {122102} (\bibinfo {year} {2009})}\BibitemShut {NoStop}%
\bibitem [{\citenamefont {Song}\ and\ \citenamefont
  {Yao}(2010)}]{Song_CPL_2010}%
  \BibitemOpen
  \bibfield  {author} {\bibinfo {author} {\bibfnamefont {C.-Y.}\ \bibnamefont
  {Song}}\ and\ \bibinfo {author} {\bibfnamefont {J.-M.}\ \bibnamefont {Yao}},\
  }\href {\doibase 10.1088/1674-1137/34/9/061} {\bibfield  {journal} {\bibinfo
  {journal} {Chin. Phys. Lett.}\ }\textbf {\bibinfo {volume} {34}},\ \bibinfo
  {pages} {1425} (\bibinfo {year} {2010})}\BibitemShut {NoStop}%
\bibitem [{\citenamefont {Song}\ \emph {et~al.}(2011)\citenamefont {Song},
  \citenamefont {Yao},\ and\ \citenamefont {Meng}}]{Song_CPL_2011}%
  \BibitemOpen
  \bibfield  {author} {\bibinfo {author} {\bibfnamefont {C.-Y.}\ \bibnamefont
  {Song}}, \bibinfo {author} {\bibfnamefont {J.-M.}\ \bibnamefont {Yao}}, \
  and\ \bibinfo {author} {\bibfnamefont {J.}~\bibnamefont {Meng}},\ }\href
  {\doibase 10.1088/0256-307x/28/9/092101} {\bibfield  {journal} {\bibinfo
  {journal} {Chin. Phys. Lett.}\ }\textbf {\bibinfo {volume} {28}},\ \bibinfo
  {pages} {092101} (\bibinfo {year} {2011})}\BibitemShut {NoStop}%
\bibitem [{\citenamefont {Sun}\ \emph {et~al.}(2017)\citenamefont {Sun},
  \citenamefont {Lu},\ and\ \citenamefont {Zhang}}]{Sun_PRC_2017}%
  \BibitemOpen
  \bibfield  {author} {\bibinfo {author} {\bibfnamefont {T.-T.}\ \bibnamefont
  {Sun}}, \bibinfo {author} {\bibfnamefont {W.-L.}\ \bibnamefont {Lu}}, \ and\
  \bibinfo {author} {\bibfnamefont {S.-S.}\ \bibnamefont {Zhang}},\ }\href
  {\doibase 10.1103/PhysRevC.96.044312} {\bibfield  {journal} {\bibinfo
  {journal} {Phys. Rev. C}\ }\textbf {\bibinfo {volume} {96}},\ \bibinfo
  {pages} {044312} (\bibinfo {year} {2017})}\BibitemShut {NoStop}%
\bibitem [{\citenamefont {Lu}\ \emph {et~al.}(2017)\citenamefont {Lu},
  \citenamefont {Liu}, \citenamefont {Ren}, \citenamefont {Zhang},\ and\
  \citenamefont {Sun}}]{Lu_JPG_2017}%
  \BibitemOpen
  \bibfield  {author} {\bibinfo {author} {\bibfnamefont {W.-L.}\ \bibnamefont
  {Lu}}, \bibinfo {author} {\bibfnamefont {Z.-X.}\ \bibnamefont {Liu}},
  \bibinfo {author} {\bibfnamefont {S.-H.}\ \bibnamefont {Ren}}, \bibinfo
  {author} {\bibfnamefont {W.}~\bibnamefont {Zhang}}, \ and\ \bibinfo {author}
  {\bibfnamefont {T.-T.}\ \bibnamefont {Sun}},\ }\href {\doibase
  10.1088/1361-6471/aa8e2d} {\bibfield  {journal} {\bibinfo  {journal} {J.
  Phys. G: Nucl. Phys.}\ }\textbf {\bibinfo {volume} {44}},\ \bibinfo {pages}
  {125104} (\bibinfo {year} {2017})}\BibitemShut {NoStop}%
\bibitem [{\citenamefont {Zhou}\ \emph {et~al.}(2003)\citenamefont {Zhou},
  \citenamefont {Meng},\ and\ \citenamefont {Ring}}]{Zhou_PRL_2003}%
  \BibitemOpen
  \bibfield  {author} {\bibinfo {author} {\bibfnamefont {S.-G.}\ \bibnamefont
  {Zhou}}, \bibinfo {author} {\bibfnamefont {J.}~\bibnamefont {Meng}}, \ and\
  \bibinfo {author} {\bibfnamefont {P.}~\bibnamefont {Ring}},\ }\href {\doibase
  10.1103/PhysRevLett.91.262501} {\bibfield  {journal} {\bibinfo  {journal}
  {Phys. Rev. Lett.}\ }\textbf {\bibinfo {volume} {91}},\ \bibinfo {pages}
  {262501} (\bibinfo {year} {2003})}\BibitemShut {NoStop}%
\bibitem [{\citenamefont {Mishustin}\ \emph {et~al.}(2005)\citenamefont
  {Mishustin}, \citenamefont {Satarov}, \citenamefont {B\"urvenich},
  \citenamefont {St\"ocker},\ and\ \citenamefont
  {Greiner}}]{Mishustin_PRC_2005}%
  \BibitemOpen
  \bibfield  {author} {\bibinfo {author} {\bibfnamefont {I.~N.}\ \bibnamefont
  {Mishustin}}, \bibinfo {author} {\bibfnamefont {L.~M.}\ \bibnamefont
  {Satarov}}, \bibinfo {author} {\bibfnamefont {T.~J.}\ \bibnamefont
  {B\"urvenich}}, \bibinfo {author} {\bibfnamefont {H.}~\bibnamefont
  {St\"ocker}}, \ and\ \bibinfo {author} {\bibfnamefont {W.}~\bibnamefont
  {Greiner}},\ }\href {\doibase 10.1103/PhysRevC.71.035201} {\bibfield
  {journal} {\bibinfo  {journal} {Phys. Rev. C}\ }\textbf {\bibinfo {volume}
  {71}},\ \bibinfo {pages} {035201} (\bibinfo {year} {2005})}\BibitemShut
  {NoStop}%
\bibitem [{\citenamefont {He}\ \emph {et~al.}(2006)\citenamefont {He},
  \citenamefont {Zhou}, \citenamefont {Meng}, \citenamefont {Zhao},\ and\
  \citenamefont {Scheid}}]{He_EPJA_2006}%
  \BibitemOpen
  \bibfield  {author} {\bibinfo {author} {\bibfnamefont {X.~T.}\ \bibnamefont
  {He}}, \bibinfo {author} {\bibfnamefont {S.~G.}\ \bibnamefont {Zhou}},
  \bibinfo {author} {\bibfnamefont {J.}~\bibnamefont {Meng}}, \bibinfo {author}
  {\bibfnamefont {E.~G.}\ \bibnamefont {Zhao}}, \ and\ \bibinfo {author}
  {\bibfnamefont {W.}~\bibnamefont {Scheid}},\ }\href {\doibase
  10.1140/epja/i2006-10066-0} {\bibfield  {journal} {\bibinfo  {journal} {Eur.
  Phys. J. A}\ }\textbf {\bibinfo {volume} {28}},\ \bibinfo {pages} {265}
  (\bibinfo {year} {2006})}\BibitemShut {NoStop}%
\bibitem [{\citenamefont {Liang}\ \emph {et~al.}(2010)\citenamefont {Liang},
  \citenamefont {Long}, \citenamefont {Meng},\ and\ \citenamefont
  {Van~Giai}}]{Liang_EPJA_2010}%
  \BibitemOpen
  \bibfield  {author} {\bibinfo {author} {\bibfnamefont {H.}~\bibnamefont
  {Liang}}, \bibinfo {author} {\bibfnamefont {W.-H.}\ \bibnamefont {Long}},
  \bibinfo {author} {\bibfnamefont {J.}~\bibnamefont {Meng}}, \ and\ \bibinfo
  {author} {\bibfnamefont {N.}~\bibnamefont {Van~Giai}},\ }\href {\doibase
  10.1140/epja/i2010-10938-6} {\bibfield  {journal} {\bibinfo  {journal} {Eur.
  Phys. J. A}\ }\textbf {\bibinfo {volume} {44}},\ \bibinfo {pages} {119}
  (\bibinfo {year} {2010})}\BibitemShut {NoStop}%
\bibitem [{\citenamefont {Guo}\ \emph {et~al.}(2005)\citenamefont {Guo},
  \citenamefont {Wang},\ and\ \citenamefont {Fang}}]{Guo_PRC_2005}%
  \BibitemOpen
  \bibfield  {author} {\bibinfo {author} {\bibfnamefont {J.-Y.}\ \bibnamefont
  {Guo}}, \bibinfo {author} {\bibfnamefont {R.-D.}\ \bibnamefont {Wang}}, \
  and\ \bibinfo {author} {\bibfnamefont {X.-Z.}\ \bibnamefont {Fang}},\ }\href
  {\doibase 10.1103/PhysRevC.72.054319} {\bibfield  {journal} {\bibinfo
  {journal} {Phys. Rev. C}\ }\textbf {\bibinfo {volume} {72}},\ \bibinfo
  {pages} {054319} (\bibinfo {year} {2005})}\BibitemShut {NoStop}%
\bibitem [{\citenamefont {Guo}\ and\ \citenamefont
  {Fang}(2006)}]{Guo_PRC_2006}%
  \BibitemOpen
  \bibfield  {author} {\bibinfo {author} {\bibfnamefont {J.~Y.}\ \bibnamefont
  {Guo}}\ and\ \bibinfo {author} {\bibfnamefont {X.~Z.}\ \bibnamefont {Fang}},\
  }\href {\doibase 10.1103/PhysRevC.74.024320} {\bibfield  {journal} {\bibinfo
  {journal} {Phys. Rev. C}\ }\textbf {\bibinfo {volume} {74}},\ \bibinfo
  {pages} {024320} (\bibinfo {year} {2006})}\BibitemShut {NoStop}%
\bibitem [{\citenamefont {Zhang}\ \emph {et~al.}(2007)\citenamefont {Zhang},
  \citenamefont {Sun},\ and\ \citenamefont {Zhou}}]{Zhang_CPL_2007}%
  \BibitemOpen
  \bibfield  {author} {\bibinfo {author} {\bibfnamefont {S.-S.}\ \bibnamefont
  {Zhang}}, \bibinfo {author} {\bibfnamefont {B.-H.}\ \bibnamefont {Sun}}, \
  and\ \bibinfo {author} {\bibfnamefont {S.-G.}\ \bibnamefont {Zhou}},\ }\href
  {\doibase 10.1088/0256-307x/24/5/020} {\bibfield  {journal} {\bibinfo
  {journal} {Chin. Phys. Lett.}\ }\textbf {\bibinfo {volume} {24}},\ \bibinfo
  {pages} {1199} (\bibinfo {year} {2007})}\BibitemShut {NoStop}%
\bibitem [{\citenamefont {Lu}\ \emph {et~al.}(2012)\citenamefont {Lu},
  \citenamefont {Zhao},\ and\ \citenamefont {Zhou}}]{Lu_PRL_2012}%
  \BibitemOpen
  \bibfield  {author} {\bibinfo {author} {\bibfnamefont {B.-N.}\ \bibnamefont
  {Lu}}, \bibinfo {author} {\bibfnamefont {E.-G.}\ \bibnamefont {Zhao}}, \ and\
  \bibinfo {author} {\bibfnamefont {S.-G.}\ \bibnamefont {Zhou}},\ }\href
  {\doibase 10.1103/PhysRevLett.109.072501} {\bibfield  {journal} {\bibinfo
  {journal} {Phys. Rev. Lett.}\ }\textbf {\bibinfo {volume} {109}},\ \bibinfo
  {pages} {072501} (\bibinfo {year} {2012})}\BibitemShut {NoStop}%
\bibitem [{\citenamefont {Lu}\ \emph {et~al.}(2013)\citenamefont {Lu},
  \citenamefont {Zhao},\ and\ \citenamefont {Zhou}}]{Lu_PRC_2013}%
  \BibitemOpen
  \bibfield  {author} {\bibinfo {author} {\bibfnamefont {B.-N.}\ \bibnamefont
  {Lu}}, \bibinfo {author} {\bibfnamefont {E.-G.}\ \bibnamefont {Zhao}}, \ and\
  \bibinfo {author} {\bibfnamefont {S.-G.}\ \bibnamefont {Zhou}},\ }\href
  {\doibase 10.1103/PhysRevC.88.024323} {\bibfield  {journal} {\bibinfo
  {journal} {Phys. Rev. C}\ }\textbf {\bibinfo {volume} {88}},\ \bibinfo
  {pages} {024323} (\bibinfo {year} {2013})}\BibitemShut {NoStop}%
\bibitem [{\citenamefont {Liu}\ \emph {et~al.}(2013)\citenamefont {Liu},
  \citenamefont {Niu},\ and\ \citenamefont {Guo}}]{Liu_PRA_2013}%
  \BibitemOpen
  \bibfield  {author} {\bibinfo {author} {\bibfnamefont {Q.}~\bibnamefont
  {Liu}}, \bibinfo {author} {\bibfnamefont {Z.-M.}\ \bibnamefont {Niu}}, \ and\
  \bibinfo {author} {\bibfnamefont {J.-Y.}\ \bibnamefont {Guo}},\ }\href
  {\doibase 10.1103/PhysRevA.87.052122} {\bibfield  {journal} {\bibinfo
  {journal} {Phys. Rev. A}\ }\textbf {\bibinfo {volume} {87}},\ \bibinfo
  {pages} {052122} (\bibinfo {year} {2013})}\BibitemShut {NoStop}%
\bibitem [{\citenamefont {Sun}\ \emph
    {et~al.}(2019{\natexlab{a}})\citenamefont
  {Sun}, \citenamefont {Lu}, \citenamefont {Qian},\ and\ \citenamefont
  {Li}}]{Sun_PRC_2019_034310}%
  \BibitemOpen
  \bibfield  {author} {\bibinfo {author} {\bibfnamefont {T.-T.}\ \bibnamefont
  {Sun}}, \bibinfo {author} {\bibfnamefont {W.-L.}\ \bibnamefont {Lu}},
  \bibinfo {author} {\bibfnamefont {L.}~\bibnamefont {Qian}}, \ and\ \bibinfo
  {author} {\bibfnamefont {Y.-X.}\ \bibnamefont {Li}},\ }\href {\doibase
  10.1103/PhysRevC.99.034310} {\bibfield  {journal} {\bibinfo  {journal} {Phys.
  Rev. C}\ }\textbf {\bibinfo {volume} {99}},\ \bibinfo {pages} {034310}
  (\bibinfo {year} {2019}{\natexlab{a}})}\BibitemShut {NoStop}%
\bibitem [{\citenamefont {Shi}\ \emph {et~al.}(2020)\citenamefont {Shi},
  \citenamefont {Liu}, \citenamefont {Guo},\ and\ \citenamefont
  {Ren}}]{ShiXX_PLB_2020}%
  \BibitemOpen
  \bibfield  {author} {\bibinfo {author} {\bibfnamefont {X.-X.}\ \bibnamefont
  {Shi}}, \bibinfo {author} {\bibfnamefont {Q.}~\bibnamefont {Liu}}, \bibinfo
  {author} {\bibfnamefont {J.-Y.}\ \bibnamefont {Guo}}, \ and\ \bibinfo
  {author} {\bibfnamefont {Z.-Z.}\ \bibnamefont {Ren}},\ }\href {\doibase
  https://doi.org/10.1016/j.physletb.2019.135174} {\bibfield  {journal}
  {\bibinfo  {journal} {Phys. Lett. B}\ }\textbf {\bibinfo {volume} {801}},\
  \bibinfo {pages} {135174} (\bibinfo {year} {2020})}\BibitemShut {NoStop}%
\bibitem [{\citenamefont {Liu}\ \emph {et~al.}(2022)\citenamefont {Liu},
  \citenamefont {Zhang},\ and\ \citenamefont {Guo}}]{Liu_PLB_2022}%
  \BibitemOpen
  \bibfield  {author} {\bibinfo {author} {\bibfnamefont {Q.}~\bibnamefont
  {Liu}}, \bibinfo {author} {\bibfnamefont {Y.}~\bibnamefont {Zhang}}, \ and\
  \bibinfo {author} {\bibfnamefont {J.-Y.}\ \bibnamefont {Guo}},\ }\href
  {\doibase https://doi.org/10.1016/j.physletb.2021.136829} {\bibfield
  {journal} {\bibinfo  {journal} {Phys. Lett. B}\ }\textbf {\bibinfo {volume}
  {824}},\ \bibinfo {pages} {136829} (\bibinfo {year} {2022})}\BibitemShut
  {NoStop}%
\bibitem [{\citenamefont {Liang}\ \emph {et~al.}(2011)\citenamefont {Liang},
  \citenamefont {Zhao}, \citenamefont {Zhang}, \citenamefont {Meng},\ and\
  \citenamefont {Giai}}]{Liang_PRC_2011}%
  \BibitemOpen
  \bibfield  {author} {\bibinfo {author} {\bibfnamefont {H.}~\bibnamefont
  {Liang}}, \bibinfo {author} {\bibfnamefont {P.}~\bibnamefont {Zhao}},
  \bibinfo {author} {\bibfnamefont {Y.}~\bibnamefont {Zhang}}, \bibinfo
  {author} {\bibfnamefont {J.}~\bibnamefont {Meng}}, \ and\ \bibinfo {author}
  {\bibfnamefont {N.~V.}\ \bibnamefont {Giai}},\ }\href {\doibase
  10.1103/PhysRevC.83.041301} {\bibfield  {journal} {\bibinfo  {journal} {Phys.
  Rev. C}\ }\textbf {\bibinfo {volume} {83}},\ \bibinfo {pages} {041301}
  (\bibinfo {year} {2011})}\BibitemShut {NoStop}%
\bibitem [{\citenamefont {Li}\ \emph {et~al.}(2011)\citenamefont {Li},
  \citenamefont {Zhao},\ and\ \citenamefont {Liang}}]{Li_CPC_2011}%
  \BibitemOpen
  \bibfield  {author} {\bibinfo {author} {\bibfnamefont {F.-Q.}\ \bibnamefont
  {Li}}, \bibinfo {author} {\bibfnamefont {P.-W.}\ \bibnamefont {Zhao}}, \ and\
  \bibinfo {author} {\bibfnamefont {H.-Z.}\ \bibnamefont {Liang}},\ }\href
  {\doibase 10.1088/1674-1137/35/9/007} {\bibfield  {journal} {\bibinfo
  {journal} {Chin. Phys. C}\ }\textbf {\bibinfo {volume} {35}},\ \bibinfo
  {pages} {825} (\bibinfo {year} {2011})}\BibitemShut {NoStop}%
\bibitem [{\citenamefont {Liang}\ \emph {et~al.}(2013)\citenamefont {Liang},
  \citenamefont {Shen}, \citenamefont {Zhao},\ and\ \citenamefont
  {Meng}}]{Liang_PRC_2013}%
  \BibitemOpen
  \bibfield  {author} {\bibinfo {author} {\bibfnamefont {H.}~\bibnamefont
  {Liang}}, \bibinfo {author} {\bibfnamefont {S.}~\bibnamefont {Shen}},
  \bibinfo {author} {\bibfnamefont {P.}~\bibnamefont {Zhao}}, \ and\ \bibinfo
  {author} {\bibfnamefont {J.}~\bibnamefont {Meng}},\ }\href {\doibase
  10.1103/PhysRevC.87.014334} {\bibfield  {journal} {\bibinfo  {journal} {Phys.
  Rev. C}\ }\textbf {\bibinfo {volume} {87}},\ \bibinfo {pages} {014334}
  (\bibinfo {year} {2013})}\BibitemShut {NoStop}%
\bibitem [{\citenamefont {Shen}\ \emph {et~al.}(2013)\citenamefont {Shen},
  \citenamefont {Liang}, \citenamefont {Zhao}, \citenamefont {Zhang},\ and\
  \citenamefont {Meng}}]{Shen_PRC_2013}%
  \BibitemOpen
  \bibfield  {author} {\bibinfo {author} {\bibfnamefont {S.}~\bibnamefont
  {Shen}}, \bibinfo {author} {\bibfnamefont {H.}~\bibnamefont {Liang}},
  \bibinfo {author} {\bibfnamefont {P.}~\bibnamefont {Zhao}}, \bibinfo {author}
  {\bibfnamefont {S.}~\bibnamefont {Zhang}}, \ and\ \bibinfo {author}
  {\bibfnamefont {J.}~\bibnamefont {Meng}},\ }\href {\doibase
  10.1103/PhysRevC.88.024311} {\bibfield  {journal} {\bibinfo  {journal} {Phys.
  Rev. C}\ }\textbf {\bibinfo {volume} {88}},\ \bibinfo {pages} {024311}
  (\bibinfo {year} {2013})}\BibitemShut {NoStop}%
\bibitem [{\citenamefont {Mottelson}(1991)}]{Mottelson_NPA_1991}%
  \BibitemOpen
  \bibfield  {author} {\bibinfo {author} {\bibfnamefont {B.}~\bibnamefont
  {Mottelson}},\ }\href {\doibase 10.1016/0375-9474(91)90048-B} {\bibfield
  {journal} {\bibinfo  {journal} {Nucl. Phys. A}\ }\textbf {\bibinfo {volume}
  {522}},\ \bibinfo {pages} {1} (\bibinfo {year} {1991})}\BibitemShut {NoStop}%
\bibitem [{\citenamefont {Sugawara-Tanabe}\ and\ \citenamefont
  {Arima}(1998)}]{Sugawara-Tanabe_PRC_1998}%
  \BibitemOpen
  \bibfield  {author} {\bibinfo {author} {\bibfnamefont {K.}~\bibnamefont
  {Sugawara-Tanabe}}\ and\ \bibinfo {author} {\bibfnamefont {A.}~\bibnamefont
  {Arima}},\ }\href {\doibase 10.1103/PhysRevC.58.R3065} {\bibfield  {journal}
  {\bibinfo  {journal} {Phys. Rev. C}\ }\textbf {\bibinfo {volume} {58}},\
  \bibinfo {pages} {R3065} (\bibinfo {year} {1998})}\BibitemShut {NoStop}%
\bibitem [{\citenamefont {Sugawara-Tanabe}\ \emph {et~al.}(2002)\citenamefont
  {Sugawara-Tanabe}, \citenamefont {Yamaji},\ and\ \citenamefont
  {Arima}}]{Sugawara-Tanabe_PRC_2002}%
  \BibitemOpen
  \bibfield  {author} {\bibinfo {author} {\bibfnamefont {K.}~\bibnamefont
  {Sugawara-Tanabe}}, \bibinfo {author} {\bibfnamefont {S.}~\bibnamefont
  {Yamaji}}, \ and\ \bibinfo {author} {\bibfnamefont {A.}~\bibnamefont
  {Arima}},\ }\href {\doibase 10.1103/PhysRevC.65.054313} {\bibfield  {journal}
  {\bibinfo  {journal} {Phys. Rev. C}\ }\textbf {\bibinfo {volume} {65}},\
  \bibinfo {pages} {054313} (\bibinfo {year} {2002})}\BibitemShut {NoStop}%
\bibitem [{\citenamefont {Sugawara-Tanabe}\ \emph {et~al.}(2000)\citenamefont
  {Sugawara-Tanabe}, \citenamefont {Yamaji},\ and\ \citenamefont
  {Arima}}]{Sugawara-Tanabe_PRC_2000}%
  \BibitemOpen
  \bibfield  {author} {\bibinfo {author} {\bibfnamefont {K.}~\bibnamefont
  {Sugawara-Tanabe}}, \bibinfo {author} {\bibfnamefont {S.}~\bibnamefont
  {Yamaji}}, \ and\ \bibinfo {author} {\bibfnamefont {A.}~\bibnamefont
  {Arima}},\ }\href {\doibase 10.1103/PhysRevC.62.054307} {\bibfield  {journal}
  {\bibinfo  {journal} {Phys. Rev. C}\ }\textbf {\bibinfo {volume} {62}},\
  \bibinfo {pages} {054307} (\bibinfo {year} {2000})}\BibitemShut {NoStop}%
\bibitem [{\citenamefont {Lalazissis}\ \emph {et~al.}(1998)\citenamefont
  {Lalazissis}, \citenamefont {Gambhir}, \citenamefont {Maharana},
  \citenamefont {Warke},\ and\ \citenamefont {Ring}}]{Lalazissis_PRC_1998}%
  \BibitemOpen
  \bibfield  {author} {\bibinfo {author} {\bibfnamefont {G.~A.}\ \bibnamefont
  {Lalazissis}}, \bibinfo {author} {\bibfnamefont {Y.~K.}\ \bibnamefont
  {Gambhir}}, \bibinfo {author} {\bibfnamefont {J.~P.}\ \bibnamefont
  {Maharana}}, \bibinfo {author} {\bibfnamefont {C.~S.}\ \bibnamefont {Warke}},
  \ and\ \bibinfo {author} {\bibfnamefont {P.}~\bibnamefont {Ring}},\ }\href
  {\doibase 10.1103/PhysRevC.58.R45} {\bibfield  {journal} {\bibinfo  {journal}
  {Phys. Rev. C}\ }\textbf {\bibinfo {volume} {58}},\ \bibinfo {pages} {R45}
  (\bibinfo {year} {1998})}\BibitemShut {NoStop}%
\bibitem [{\citenamefont {Ginocchio}\ \emph {et~al.}(2004)\citenamefont
  {Ginocchio}, \citenamefont {Leviatan}, \citenamefont {Meng},\ and\
  \citenamefont {Zhou}}]{PRC2004Ginocchio_69_034303}%
  \BibitemOpen
  \bibfield  {author} {\bibinfo {author} {\bibfnamefont {J.~N.}\ \bibnamefont
  {Ginocchio}}, \bibinfo {author} {\bibfnamefont {A.}~\bibnamefont {Leviatan}},
  \bibinfo {author} {\bibfnamefont {J.}~\bibnamefont {Meng}}, \ and\ \bibinfo
  {author} {\bibfnamefont {S.-G.}\ \bibnamefont {Zhou}},\ }\href {\doibase
  10.1103/PhysRevC.69.034303} {\bibfield  {journal} {\bibinfo  {journal} {Phys.
  Rev. C}\ }\textbf {\bibinfo {volume} {69}},\ \bibinfo {pages} {034303}
  (\bibinfo {year} {2004})}\BibitemShut {NoStop}%
\bibitem [{\citenamefont {Guo}\ \emph {et~al.}(2014)\citenamefont {Guo},
  \citenamefont {Chen}, \citenamefont {Niu}, \citenamefont {Li},\ and\
  \citenamefont {Liu}}]{Guo_PRL_2014}%
  \BibitemOpen
  \bibfield  {author} {\bibinfo {author} {\bibfnamefont {J.~Y.}\ \bibnamefont
  {Guo}}, \bibinfo {author} {\bibfnamefont {S.~W.}\ \bibnamefont {Chen}},
  \bibinfo {author} {\bibfnamefont {Z.~M.}\ \bibnamefont {Niu}}, \bibinfo
  {author} {\bibfnamefont {D.~P.}\ \bibnamefont {Li}}, \ and\ \bibinfo {author}
  {\bibfnamefont {Q.}~\bibnamefont {Liu}},\ }\href {\doibase
  10.1103/PhysRevLett.112.062502} {\bibfield  {journal} {\bibinfo  {journal}
  {Phy. Rev. Lett.}\ }\textbf {\bibinfo {volume} {112}},\ \bibinfo {pages}
  {062502} (\bibinfo {year} {2014})}\BibitemShut {NoStop}%
\bibitem [{\citenamefont {Zhang}\ \emph {et~al.}(2023)\citenamefont {Zhang},
  \citenamefont {Luo}, \citenamefont {Liu},\ and\ \citenamefont
  {Guo}}]{LiuQ_PLB_2023}%
  \BibitemOpen
  \bibfield  {author} {\bibinfo {author} {\bibfnamefont {Y.}~\bibnamefont
  {Zhang}}, \bibinfo {author} {\bibfnamefont {Y.-X.}\ \bibnamefont {Luo}},
  \bibinfo {author} {\bibfnamefont {Q.}~\bibnamefont {Liu}}, \ and\ \bibinfo
  {author} {\bibfnamefont {J.-Y.}\ \bibnamefont {Guo}},\ }\href {\doibase
  https://doi.org/10.1016/j.physletb.2023.137716} {\bibfield  {journal}
  {\bibinfo  {journal} {Phys. Lett. B}\ }\textbf {\bibinfo {volume} {838}},\
  \bibinfo {pages} {137716} (\bibinfo {year} {2023})}\BibitemShut {NoStop}%
\bibitem [{\citenamefont {Sun}\ \emph {et~al.}(2014)\citenamefont {Sun},
  \citenamefont {Zhang}, \citenamefont {Zhang}, \citenamefont {Hu},\ and\
  \citenamefont {Meng}}]{Sun_PRC_2014}%
  \BibitemOpen
  \bibfield  {author} {\bibinfo {author} {\bibfnamefont {T.~T.}\ \bibnamefont
  {Sun}}, \bibinfo {author} {\bibfnamefont {S.~Q.}\ \bibnamefont {Zhang}},
  \bibinfo {author} {\bibfnamefont {Y.}~\bibnamefont {Zhang}}, \bibinfo
  {author} {\bibfnamefont {J.~N.}\ \bibnamefont {Hu}}, \ and\ \bibinfo {author}
  {\bibfnamefont {J.}~\bibnamefont {Meng}},\ }\href {\doibase
  10.1103/PhysRevC.90.054321} {\bibfield  {journal} {\bibinfo  {journal} {Phys.
  Rev. C}\ }\textbf {\bibinfo {volume} {90}},\ \bibinfo {pages} {054321}
  (\bibinfo {year} {2014})}\BibitemShut {NoStop}%
\bibitem [{\citenamefont {Sun}\ \emph {et~al.}(2016)\citenamefont {Sun},
  \citenamefont {Niu},\ and\ \citenamefont {Zhang}}]{TTSun_JPG_2016}%
  \BibitemOpen
  \bibfield  {author} {\bibinfo {author} {\bibfnamefont {T.-T.}\ \bibnamefont
  {Sun}}, \bibinfo {author} {\bibfnamefont {Z.-M.}\ \bibnamefont {Niu}}, \ and\
  \bibinfo {author} {\bibfnamefont {S.-Q.}\ \bibnamefont {Zhang}},\ }\href
  {\doibase 10.1088/0954-3899/43/4/045107} {\bibfield  {journal} {\bibinfo
  {journal} {J. Phys. G: Nucl. Phys.}\ }\textbf {\bibinfo {volume} {43}},\
  \bibinfo {pages} {045107} (\bibinfo {year} {2016})}\BibitemShut {NoStop}%
\bibitem [{\citenamefont {Ren}\ \emph {et~al.}(2017)\citenamefont {Ren},
  \citenamefont {Sun},\ and\ \citenamefont {Zhang}}]{SHRen_PRC_2017}%
  \BibitemOpen
  \bibfield  {author} {\bibinfo {author} {\bibfnamefont {S.-H.}\ \bibnamefont
  {Ren}}, \bibinfo {author} {\bibfnamefont {T.-T.}\ \bibnamefont {Sun}}, \ and\
  \bibinfo {author} {\bibfnamefont {W.}~\bibnamefont {Zhang}},\ }\href
  {\doibase 10.1103/PhysRevC.95.054318} {\bibfield  {journal} {\bibinfo
  {journal} {Phys. Rev. C}\ }\textbf {\bibinfo {volume} {95}},\ \bibinfo
  {pages} {054318} (\bibinfo {year} {2017})}\BibitemShut {NoStop}%
\bibitem [{\citenamefont {Qu}\ and\ \citenamefont {Zhang}(2019)}]{Qu_PRC_2019}%
  \BibitemOpen
  \bibfield  {author} {\bibinfo {author} {\bibfnamefont {X.~Y.}\ \bibnamefont
  {Qu}}\ and\ \bibinfo {author} {\bibfnamefont {Y.}~\bibnamefont {Zhang}},\
  }\href {\doibase 10.1103/PhysRevC.99.014314} {\bibfield  {journal} {\bibinfo
  {journal} {Phys. Rev. C}\ }\textbf {\bibinfo {volume} {99}},\ \bibinfo
  {pages} {014314} (\bibinfo {year} {2019})}\BibitemShut {NoStop}%
\bibitem [{\citenamefont {Qu}\ \emph {et~al.}(2022)\citenamefont {Qu},
  \citenamefont {Tong},\ and\ \citenamefont {Zhang}}]{Qu_PRC_2022}%
  \BibitemOpen
  \bibfield  {author} {\bibinfo {author} {\bibfnamefont {X.~Y.}\ \bibnamefont
  {Qu}}, \bibinfo {author} {\bibfnamefont {H.}~\bibnamefont {Tong}}, \ and\
  \bibinfo {author} {\bibfnamefont {S.~Q.}\ \bibnamefont {Zhang}},\ }\href
  {\doibase 10.1103/PhysRevC.105.014326} {\bibfield  {journal} {\bibinfo
  {journal} {Phys. Rev. C}\ }\textbf {\bibinfo {volume} {105}},\ \bibinfo
  {pages} {014326} (\bibinfo {year} {2022})}\BibitemShut {NoStop}%
\bibitem [{\citenamefont {Oba}\ and\ \citenamefont
  {Matsuo}(2009)}]{Oba_PRC_2009}%
  \BibitemOpen
  \bibfield  {author} {\bibinfo {author} {\bibfnamefont {H.}~\bibnamefont
  {Oba}}\ and\ \bibinfo {author} {\bibfnamefont {M.}~\bibnamefont {Matsuo}},\
  }\href {\doibase 10.1103/PhysRevC.80.024301} {\bibfield  {journal} {\bibinfo
  {journal} {Phys. Rev. C}\ }\textbf {\bibinfo {volume} {80}},\ \bibinfo
  {pages} {024301} (\bibinfo {year} {2009})}\BibitemShut {NoStop}%
\bibitem [{\citenamefont {Zhang}\ \emph {et~al.}(2011)\citenamefont {Zhang},
  \citenamefont {Matsuo},\ and\ \citenamefont {Meng}}]{Zhang_PRC_2011}%
  \BibitemOpen
  \bibfield  {author} {\bibinfo {author} {\bibfnamefont {Y.}~\bibnamefont
  {Zhang}}, \bibinfo {author} {\bibfnamefont {M.}~\bibnamefont {Matsuo}}, \
  and\ \bibinfo {author} {\bibfnamefont {J.}~\bibnamefont {Meng}},\ }\href
  {\doibase 10.1103/PhysRevC.83.054301} {\bibfield  {journal} {\bibinfo
  {journal} {Phys. Rev. C}\ }\textbf {\bibinfo {volume} {83}},\ \bibinfo
  {pages} {054301} (\bibinfo {year} {2011})}\BibitemShut {NoStop}%
\bibitem [{\citenamefont {Sun}(2016)}]{TTSun_Sci_2016}%
  \BibitemOpen
  \bibfield  {author} {\bibinfo {author} {\bibfnamefont {T.-T.}\ \bibnamefont
  {Sun}},\ }\href {\doibase doi:10.1360/SSPMA2015-00371} {\bibfield  {journal}
  {\bibinfo  {journal} {Sci. Sin.-Phys. Mech. Astron.}\ }\textbf {\bibinfo
  {volume} {46}},\ \bibinfo {pages} {12006} (\bibinfo {year}
  {2016})}\BibitemShut {NoStop}%
\bibitem [{\citenamefont {Sun}\ \emph
    {et~al.}(2019{\natexlab{b}})\citenamefont
  {Sun}, \citenamefont {Liu}, \citenamefont {Qian}, \citenamefont {Wang},\ and\
  \citenamefont {Zhang}}]{Sun_PRC_2019_054316}%
  \BibitemOpen
  \bibfield  {author} {\bibinfo {author} {\bibfnamefont {T.-T.}\ \bibnamefont
  {Sun}}, \bibinfo {author} {\bibfnamefont {Z.-X.}\ \bibnamefont {Liu}},
  \bibinfo {author} {\bibfnamefont {L.}~\bibnamefont {Qian}}, \bibinfo {author}
  {\bibfnamefont {B.}~\bibnamefont {Wang}}, \ and\ \bibinfo {author}
  {\bibfnamefont {W.}~\bibnamefont {Zhang}},\ }\href {\doibase
  10.1103/PhysRevC.99.054316} {\bibfield  {journal} {\bibinfo  {journal} {Phys.
  Rev. C}\ }\textbf {\bibinfo {volume} {99}},\ \bibinfo {pages} {054316}
  (\bibinfo {year} {2019}{\natexlab{b}})}\BibitemShut {NoStop}%
\bibitem [{\citenamefont {Matsuo}(2001)}]{Matsuo_NPA_2001}%
  \BibitemOpen
  \bibfield  {author} {\bibinfo {author} {\bibfnamefont {M.}~\bibnamefont
  {Matsuo}},\ }\href {\doibase https://doi.org/10.1016/S0375-9474(01)01133-2}
  {\bibfield  {journal} {\bibinfo  {journal} {Nucl. Phys. A}\ }\textbf
  {\bibinfo {volume} {696}},\ \bibinfo {pages} {371} (\bibinfo {year}
  {2001})}\BibitemShut {NoStop}%
\bibitem [{\citenamefont {Matsuo}\ \emph {et~al.}(2005)\citenamefont {Matsuo},
  \citenamefont {Mizuyama},\ and\ \citenamefont {Serizawa}}]{Matsuo_PRC_2005}%
  \BibitemOpen
  \bibfield  {author} {\bibinfo {author} {\bibfnamefont {M.}~\bibnamefont
  {Matsuo}}, \bibinfo {author} {\bibfnamefont {K.}~\bibnamefont {Mizuyama}}, \
  and\ \bibinfo {author} {\bibfnamefont {Y.}~\bibnamefont {Serizawa}},\ }\href
  {\doibase 10.1103/PhysRevC.71.064326} {\bibfield  {journal} {\bibinfo
  {journal} {Phys. Rev. C}\ }\textbf {\bibinfo {volume} {71}},\ \bibinfo
  {pages} {064326} (\bibinfo {year} {2005})}\BibitemShut {NoStop}%
\bibitem [{\citenamefont {Matsuo}\ and\ \citenamefont
  {Serizawa}(2010)}]{Matsuo_PRC_2010}%
  \BibitemOpen
  \bibfield  {author} {\bibinfo {author} {\bibfnamefont {M.}~\bibnamefont
  {Matsuo}}\ and\ \bibinfo {author} {\bibfnamefont {Y.}~\bibnamefont
  {Serizawa}},\ }\href {\doibase 10.1103/PhysRevC.82.024318} {\bibfield
  {journal} {\bibinfo  {journal} {Phys. Rev. C}\ }\textbf {\bibinfo {volume}
  {82}},\ \bibinfo {pages} {024318} (\bibinfo {year} {2010})}\BibitemShut
  {NoStop}%
\bibitem [{\citenamefont {Shimoyama}\ and\ \citenamefont
  {Matsuo}(2011)}]{Shimoyama_PRC_2011}%
  \BibitemOpen
  \bibfield  {author} {\bibinfo {author} {\bibfnamefont {H.}~\bibnamefont
  {Shimoyama}}\ and\ \bibinfo {author} {\bibfnamefont {M.}~\bibnamefont
  {Matsuo}},\ }\href {\doibase 10.1103/PhysRevC.84.044317} {\bibfield
  {journal} {\bibinfo  {journal} {Phys. Rev. C}\ }\textbf {\bibinfo {volume}
  {84}},\ \bibinfo {pages} {044317} (\bibinfo {year} {2011})}\BibitemShut
  {NoStop}%
\bibitem [{\citenamefont {Shimoyama}\ and\ \citenamefont
  {Matsuo}(2013)}]{Shimoyama_PRC_2013}%
  \BibitemOpen
  \bibfield  {author} {\bibinfo {author} {\bibfnamefont {H.}~\bibnamefont
  {Shimoyama}}\ and\ \bibinfo {author} {\bibfnamefont {M.}~\bibnamefont
  {Matsuo}},\ }\href {\doibase 10.1103/PhysRevC.88.054308} {\bibfield
  {journal} {\bibinfo  {journal} {Phys. Rev. C}\ }\textbf {\bibinfo {volume}
  {88}},\ \bibinfo {pages} {054308} (\bibinfo {year} {2013})}\BibitemShut
  {NoStop}%
\bibitem [{\citenamefont {Matsuo}(2015)}]{Matsuo_PRC_2015}%
  \BibitemOpen
  \bibfield  {author} {\bibinfo {author} {\bibfnamefont {M.}~\bibnamefont
  {Matsuo}},\ }\href {\doibase 10.1103/PhysRevC.91.034604} {\bibfield
  {journal} {\bibinfo  {journal} {Phys. Rev. C}\ }\textbf {\bibinfo {volume}
  {91}},\ \bibinfo {pages} {034604} (\bibinfo {year} {2015})}\BibitemShut
  {NoStop}%
\bibitem [{\citenamefont {Huo}\ \emph {et~al.}(2023)\citenamefont {Huo},
  \citenamefont {Li}, \citenamefont {Qu}, \citenamefont {Zhang},\ and\
  \citenamefont {Sun}}]{Huo_NST_2023}%
  \BibitemOpen
  \bibfield  {author} {\bibinfo {author} {\bibfnamefont {E.-B.}\ \bibnamefont
  {Huo}}, \bibinfo {author} {\bibfnamefont {K.-R.}\ \bibnamefont {Li}},
  \bibinfo {author} {\bibfnamefont {X.-Y.}\ \bibnamefont {Qu}}, \bibinfo
  {author} {\bibfnamefont {Y.}~\bibnamefont {Zhang}}, \ and\ \bibinfo {author}
  {\bibfnamefont {T.-T.}\ \bibnamefont {Sun}},\ }\href {\doibase
  https://doi.org/10.1007/s41365-023-01261-9} {\bibfield  {journal} {\bibinfo
  {journal} {Nucl. Sci. Tech.}\ }\textbf {\bibinfo {volume} {34}},\ \bibinfo
  {pages} {105} (\bibinfo {year} {2023})}\BibitemShut {NoStop}%
\bibitem [{\citenamefont {Sun}\ \emph {et~al.}(2020)\citenamefont {Sun},
  \citenamefont {Qian}, \citenamefont {Chen}, \citenamefont {Ring},\ and\
  \citenamefont {Li}}]{Sun_PRC_2020}%
  \BibitemOpen
  \bibfield  {author} {\bibinfo {author} {\bibfnamefont {T.-T.}\ \bibnamefont
  {Sun}}, \bibinfo {author} {\bibfnamefont {L.}~\bibnamefont {Qian}}, \bibinfo
  {author} {\bibfnamefont {C.}~\bibnamefont {Chen}}, \bibinfo {author}
  {\bibfnamefont {P.}~\bibnamefont {Ring}}, \ and\ \bibinfo {author}
  {\bibfnamefont {Z.~P.}\ \bibnamefont {Li}},\ }\href {\doibase
  10.1103/PhysRevC.101.014321} {\bibfield  {journal} {\bibinfo  {journal}
  {Phys. Rev. C}\ }\textbf {\bibinfo {volume} {101}},\ \bibinfo {pages}
  {014321} (\bibinfo {year} {2020})}\BibitemShut {NoStop}%
\bibitem [{\citenamefont {Chen}\ \emph {et~al.}(2020)\citenamefont {Chen},
  \citenamefont {Li}, \citenamefont {Li},\ and\ \citenamefont
  {Sun}}]{Chen_CPC_2020}%
  \BibitemOpen
  \bibfield  {author} {\bibinfo {author} {\bibfnamefont {C.}~\bibnamefont
  {Chen}}, \bibinfo {author} {\bibfnamefont {Z.~P.}\ \bibnamefont {Li}},
  \bibinfo {author} {\bibfnamefont {Y.~X.}\ \bibnamefont {Li}}, \ and\ \bibinfo
  {author} {\bibfnamefont {T.-T.}\ \bibnamefont {Sun}},\ }\href {\doibase
  10.1088/1674-1137/44/8/084105} {\bibfield  {journal} {\bibinfo  {journal}
  {Chin. Phys. C}\ }\textbf {\bibinfo {volume} {44}},\ \bibinfo {pages}
  {084105} (\bibinfo {year} {2020})}\BibitemShut {NoStop}%
\bibitem [{\citenamefont {Wang}\ and\ \citenamefont
  {Sun}(2021)}]{Wang_NST_2021}%
  \BibitemOpen
  \bibfield  {author} {\bibinfo {author} {\bibfnamefont {Y.-T.}\ \bibnamefont
  {Wang}}\ and\ \bibinfo {author} {\bibfnamefont {T.-T.}\ \bibnamefont {Sun}},\
  }\href {\doibase 10.1007/s41365-021-00884-0} {\bibfield  {journal} {\bibinfo
  {journal} {Nucl. Sci. Tech.}\ }\textbf {\bibinfo {volume} {32}},\ \bibinfo
  {pages} {46} (\bibinfo {year} {2021})}\BibitemShut {NoStop}%
\bibitem [{\citenamefont {Sun}\ \emph {et~al.}(2023)\citenamefont {Sun},
  \citenamefont {Li},\ and\ \citenamefont {Ring}}]{PLB2023TTSun}%
  \BibitemOpen
  \bibfield  {author} {\bibinfo {author} {\bibfnamefont {T.-T.}\ \bibnamefont
  {Sun}}, \bibinfo {author} {\bibfnamefont {Z.~P.}\ \bibnamefont {Li}}, \ and\
  \bibinfo {author} {\bibfnamefont {P.}~\bibnamefont {Ring}},\ }\href {\doibase 10.1016/j.physletb.2023.138320}
  {\bibfield  {journal} {\bibinfo  {journal} {Phys. Lett. B}\ }\textbf {\bibinfo {volume} {847}},\ \bibinfo
  {pages} {138320} (\bibinfo {year} {2023})}\BibitemShut {NoStop}%
\bibitem [{\citenamefont {Hamamoto}\ (2004)}]{PRC2004Hamamoto_69_041306}%
  \BibitemOpen
  \bibfield  {author} {\bibinfo {author} {\bibfnamefont {I.}~\bibnamefont
  {Hamamoto}},\ }\href {\doibase 10.1103/PhysRevC.69.041306} {\bibfield
  {journal} {\bibinfo  {journal} {Phys. Rev. C}\ }\textbf {\bibinfo {volume}
  {69}},\ \bibinfo {pages} {041306} (\bibinfo {year} {2004})}\BibitemShut
  {NoStop}%
\bibitem [{\citenamefont {Li}\ \emph {et~al.}(2020)\citenamefont {Li},
  \citenamefont {Meng}, \citenamefont {Zhang}, \citenamefont {Zhou},\ and\
  \citenamefont {Savushkin}}]{PRC2010ZPLi_81_034311}%
  \BibitemOpen
  \bibfield  {author} {\bibinfo {author} {\bibfnamefont {Z. P.}\ \bibnamefont
  {Li}}, \bibinfo {author} {\bibfnamefont {J.}~\bibnamefont {Meng}}, \bibinfo
  {author} {\bibfnamefont {Y.}~\bibnamefont {Zhang}}, \bibinfo {author}
  {\bibfnamefont {S. G.}~\bibnamefont {Zhou}}, \ and\ \bibinfo {author}
  {\bibfnamefont {L.~N.}\ \bibnamefont {Savushkin}},\ }\href {\doibase
  10.1103/PhysRevC.81.034311} {\bibfield  {journal} {\bibinfo  {journal}
  {Phys. Rev. C}\ }\textbf {\bibinfo {volume} {81}},\ \bibinfo {pages}
  {034311} (\bibinfo {year} {2010})}\BibitemShut {NoStop}%
\bibitem [{\citenamefont {Pritychenko}\ \emph {et~al.}(2016)\citenamefont
  {Pritychenko}, \citenamefont {Birch}, \citenamefont {Singh},\ and\
  \citenamefont {Horoi}}]{ADNDT2016}%
  \BibitemOpen
  \bibfield  {author} {\bibinfo {author} {\bibfnamefont {B.}~\bibnamefont
  {Pritychenko}}, \bibinfo {author} {\bibfnamefont {M.}~\bibnamefont {Birch}},
  \bibinfo {author} {\bibfnamefont {B.}~\bibnamefont {Singh}}, \ and\ \bibinfo
  {author} {\bibfnamefont {M.}~\bibnamefont {Horoi}},\ }\href {\doibase
  https://doi.org/10.1016/j.adt.2015.10.001} {\bibfield  {journal} {\bibinfo
  {journal} {At. Data Nucl. Data Tables}\ }\textbf {\bibinfo {volume} {107}},\
  \bibinfo {pages} {1} (\bibinfo {year} {2016})}\BibitemShut {NoStop}%

\end{thebibliography}

%

\end{document}